\documentclass[aps,superscriptaddress,preprint]{revtex4-1}
\usepackage{lineno,hyperref}

\usepackage{graphicx}
\usepackage{epstopdf}
\usepackage{amsmath,amssymb}
\usepackage{color}
\usepackage{dcolumn}
\usepackage{bm}
\usepackage{caption}

\begin{document}

\title{Energy transfer, pressure tensor and heating of kinetic plasma}

\author{Yan Yang}
\affiliation{State Key Laboratory for Turbulence and Complex Systems, Center for Applied Physics and Technology, College of Engineering, Peking University, Beijing 100871, P. R. China}
\affiliation{Department of Physics and Astronomy, University of Delaware, Newark, DE 19716}
\author{William H. Matthaeus}
\affiliation{Department of Physics and Astronomy, University of Delaware, Newark, DE 19716}
\author{Tulasi N. Parashar}
\affiliation{Department of Physics and Astronomy, University of Delaware, Newark, DE 19716}
\author{Colby C. Haggerty}
\affiliation{Department of Physics and Astronomy, University of Delaware, Newark, DE 19716}
\author{Vadim Roytershteyn}
\affiliation{Space Science Institute, Boulder CO, 80301}
\author{William Daughton}
\affiliation{Los Alamos National Laboratory, Los Alamos, NM 87545}
\author{Minping Wan}
\affiliation{Department of Mechanics and Aerospace Engineering, South University of Science and Technology of China, Shenzhen, Guangdong 518055, China}
\author{Yipeng Shi}
\affiliation{State Key Laboratory for Turbulence and Complex Systems, Center for Applied Physics and Technology, College of Engineering, Peking University, Beijing 100871, P. R. China}
\author{Shiyi Chen}
\affiliation{Department of Mechanics and Aerospace Engineering, South University of Science and Technology of China, Shenzhen, Guangdong 518055, China}
\affiliation{State Key Laboratory for Turbulence and Complex Systems, Center for Applied Physics and Technology, College of Engineering, Peking University, Beijing 100871, P. R. China}

\date{\today}

\begin{abstract}
Kinetic plasma turbulence cascade
spans multiple scales
ranging from macroscopic fluid flow to sub-electron scales.
Mechanisms that dissipate large scale energy,
terminate the inertial range cascade and
convert kinetic energy into heat are hotly debated.
Here we revisit these puzzles using
fully kinetic simulation.
By performing scale-dependent
spatial filtering on the Vlasov equation,
we extract information at prescribed scales and
introduce several energy transfer functions.
This approach allows highly inhomogeneous
energy cascade to be quantified as
it proceeds
down to kinetic scales.
The pressure work,
$-\left( \boldsymbol{P} \cdot \nabla \right) \cdot \boldsymbol{u}$,
can trigger a channel of the energy conversion
between fluid flow and random motions,
which is a collision-free generalization of
the viscous dissipation in collisional fluid.
Both the energy transfer and the pressure work
are strongly correlated with velocity gradients.
\end{abstract}

\maketitle

\section{Introduction}
The classical energy cascade scenario is
of great importance in explaining the
heating of corona and solar wind
{\citep{Coleman68, Tu84, Matthaeus11, Schekochihin09, Howes15}}:
In these applications, one envisions that
significant amounts of energy reside in large-scale fluctuations.
Nonlinear interactions
cause a  cascade that transports  energy
to smaller scales where dissipation occurs.
At dissipation scales kinetic
processes that absorb these energy fluxes,
produce temperature enhancements.
Inhomogeneities generated in this process
may in turn be responsible for large scale flows,
while also producing populations of energetic particles.
Many systems, including
low-collisionality astrophysical plasmas,
may be well described by fluid theory at large scales.
For plasmas, including electromagnetic fields,
this large scale description would usually be taken
to be some form  of magnetohydrodynamics (MHD).
The present paper is devoted to extending ideas
about turbulence cascade into the deep kinetic range,
so that we may develop a better understanding of
turbulence cascade in
low collsionality or even collsionless plasma.

It is reasonable to assume that
MHD remains a credible approximation for a kinetic
plasma at scales large enough to
be well separated from
kinetic effects.
For that range, scale filtered
analysis \citep{YangEA-PRE-16} shows
how the fluid flow energy cascades almost conservatively from
large to small scales, despite not being strictly
an invariant of the MHD system.
Then an important subsequent question
asks how energy transfer proceeds down to kinetic scales
as various kinetic processes come to the fore.
The present paper addresses statistical properties of energy transfer
across scales,
recognizing the possible significance of energy
cascade in explaining the heating and acceleration of the wind,
and many other properties.

In this work we will avoid
adopting familiar approaches that
rely heavily on linear theory of waves, instabilities
and damping rates \cite{Gary, SchekochihinEA08,MatthaeusEA14},
or on weak turbulence approaches
that require a leading order description in terms of linear modes.
Instead we consider the full Vlasov-Maxwell system,
and employ a filtering approach
that is familiar in hydrodynamics {\citep{Aluie11b}}
and large-eddy simulation {\citep{Chernyshov06, Petrosyan10, Martin00, Agullo01, Miesch15}}
communities but less used in kinetic plasma.
Examining filtered equations for energy transfer,
we can assess the relative importance of different transfer terms at all
scales ranging from MHD to electron scales.

The present
approach provides extensions of what fluid models tell us about
the plasma cascade.
In the context of plasma applications,
MHD simulations adopt an {\it ad hoc} model of dissipation
(e.g., viscous and resistive dissipation),
rather than engaging the
details of the small dynamics
that make up the plasma dissipation range.
The turbulence in most astrophysical contexts,
on the other hand, is typically of weak collisionality,
and frequently modeled as collisionless,
and thus collisional (viscous and resistive) dissipation
at small scales cannot emerge immediately.
While various specific
processes may contribute to conversion of energy from
fields into random degrees of freedom, for example,
wave-particle interactions (WPI)
{\citep{Markovskii06, Hollweg86, Hollweg02, Gary03, Gary08}}
and processes associated with coherent structures (CS)
{\citep{Dmitruk04, Retino07, Sundkvist07, ParasharEA11, TenBarge13, Perri12}},
are likely ingredients, but nevertheless
an explicit dissipation function cannot at this moment be defined clearly
for a collisionless system.

Lacking such an explicit form for dissipation,
Wan et al. {\citep{WanEA15}} considered a surrogate dissipation measure
related to the work done by the electromagnetic field on the
plasma particles.
Recent studies in compressible MHD turbulence {\citep{YangEA-PRE-16,YangEA-POF-17}}
demonstrated that, apart from collisional dissipation,
the pressure dilatation, $-p \nabla \cdot \boldsymbol{u}$,
can trigger an alternative channel of the conversion between kinetic and internal energy.
Accordingly, one could expect that
there might be an analogous role of the pressure tensor in
collisionless plasma.
In fact one expects pressure
to be influential in at least several ways.
On the one hand,
the pressure term in anisotropic compressible turbulence
moderates the competition and balance
between two energy redistributive processes,
i.e., return-to-isotropy {\citep{book-Pope00, Crespo05}} and
kinetic-potential (internal) energy equipartition
{\citep{Sarkar92, Miura95, Lee06, Lee16}}.
On the other hand,
the pressure tensor in kinetic plasmas
plays a very important role in the force balance equations
as well as in the generalized Ohm's law near neutral lines
{\citep{Cai97, Yin01}}.
Here we show that
the global energy exchange between fluid flow and
particles (i.e., kinetic and thermal energies),
derived from the Vlasov equation, is bridged immediately by
the collaboration of pressure tensor and strain stress (i.e., velocity gradient).
This possibly provides a new perspective on the collisionless dissipation mechanism
and on the collisionless plasma cascade in general.
The possible importance of pressure work in generating
internal energy has been brought up in Ref. {\cite{YangEA-PRL-17}} (hereafter, Paper I).
Here we extend that study
and explore this novelty in a more comprehensive and detailed way.

\section{Global energy conversion}
\label{Sec.GlobalEnergyConversion}
Standard manipulation of the Vlasov equation
yields macroscopic equations for plasma particles of
type $\alpha$ in a collisionless plasma:
\begin{eqnarray}
\partial_t \rho_{\alpha} + \nabla \cdot \left(\rho_{\alpha} \boldsymbol{u}_{\alpha}\right) &=& 0,
\label{Eq.1}\\
\partial_t \left(\rho_{\alpha} \boldsymbol{u}_{\alpha}\right) + \nabla \cdot \left(\rho_{\alpha} \boldsymbol{u}_{\alpha}\boldsymbol{u}_{\alpha} \right) &=& -\nabla \cdot \boldsymbol{P}_{\alpha} + n_{\alpha} q_{\alpha} \left(\boldsymbol{E}+{\boldsymbol{u}_{\alpha}/c \times \boldsymbol{B}}\right),
\label{Eq.2}  \\
\partial_t \mathcal{E}_{\alpha} + \nabla \cdot \left(\mathcal{E}_{\alpha} \boldsymbol{u}_{\alpha}\right) &=& - \nabla \cdot \left(\boldsymbol{P}_{\alpha} \cdot \boldsymbol{u}_{\alpha}\right) - \nabla \cdot \boldsymbol{q}_{\alpha} + n_{\alpha} q_{\alpha} \boldsymbol{E} \cdot \boldsymbol{u}_{\alpha}.
\label{Eq.3}
\end{eqnarray}
Here $\rho_{\alpha}=m_{\alpha} n_{\alpha}$ represents the mass density;
$m_{\alpha}$ is the mass of particles of species $\alpha$;
$n_{\alpha}$ is the number density;
$\boldsymbol{u}_{\alpha}$ gives the fluid flow (bulk) velocity;
$n_{\alpha} q_{\alpha}$ represents the charge density;
$\boldsymbol{P}_{\alpha}=m_{\alpha} \int{\left(\boldsymbol{v}-\boldsymbol{u}_{\alpha}\right)\left(\boldsymbol{v}-\boldsymbol{u}_{\alpha}\right) f_{\alpha} \left(\boldsymbol{x},\boldsymbol{v},t\right) d\boldsymbol{v}}$ is the pressure tensor;
$\mathcal{E}_{\alpha}=\int{{\frac{1}{2}} m_{\alpha} \boldsymbol{v}^2 f_{\alpha} \left(\boldsymbol{x},\boldsymbol{v},t\right) d\boldsymbol{v}}$ is the total (average and random) kinetic energy;
$\boldsymbol{q}_{\alpha}={\frac{1}{2}} m_\alpha \int{\left(\boldsymbol{v}-\boldsymbol{u}_\alpha\right)^2  \left(\boldsymbol{v}-\boldsymbol{u}_\alpha\right) f_\alpha \left(\boldsymbol{x},\boldsymbol{v},t\right) d\boldsymbol{v}}$ is the heat flux vector.

Decomposing the total energy $\mathcal{E}_{\alpha}$
into average and random parts facilitates
the understanding of energy converting processes.
On defining the species fluid flow energy as
$E^f_\alpha={\frac{1}{2}}\rho_\alpha \boldsymbol{u}^2_\alpha$
and the thermal (random) energy as
$E^{th}_\alpha={\frac{1}{2}} m_\alpha \int{\left(\boldsymbol{v}-\boldsymbol{u}_\alpha\right)^2 f_\alpha \left(\boldsymbol{x},\boldsymbol{v},t\right) d\boldsymbol{v}}$,
it is obvious that $\mathcal{E}_{\alpha}=E^f_\alpha+E^{th}_\alpha$.
Computing the inner product of
Eq. {\ref{Eq.2}} with
$\boldsymbol{u}_{\alpha}$ results in the fluid flow energy equation:
\begin{eqnarray}
\partial_t E^{f}_\alpha + \nabla \cdot \left( E^{f}_\alpha \boldsymbol{u}_\alpha \right) &=&-\nabla \cdot \left( \boldsymbol{P}_\alpha \cdot \boldsymbol{u}_\alpha \right)+ \left( \boldsymbol{P}_\alpha \cdot \nabla \right) \cdot \boldsymbol{u}_\alpha+ n_\alpha q_\alpha \boldsymbol{E} \cdot \boldsymbol{u}_\alpha.
\label{Eq.4}
\end{eqnarray}
Subtracting Eq. {\ref{Eq.4}} from Eq. {\ref{Eq.3}} we obtain
a time evolution equation for the random kinetic energy,
\begin{eqnarray}
\partial_t E^{th}_\alpha + \nabla \cdot \left( E^{th}_\alpha \boldsymbol{u}_\alpha \right) &=& -\left( \boldsymbol{P}_\alpha \cdot \nabla \right) \cdot \boldsymbol{u}_\alpha - \nabla \cdot \boldsymbol{q}_\alpha.
\label{Eq.5}
\end{eqnarray}
Using the Maxwell curl equations,
the equation governing  electromagnetic energy,
$E^m={\frac{1}{8\pi}}\left(\boldsymbol{B}^2+\boldsymbol{E}^2\right)$, can be written as:
\begin{eqnarray}
\partial_t E^{m} + {\frac{c}{4\pi}} \nabla \cdot \left( \boldsymbol{E} \times \boldsymbol{B} \right) &=& -\boldsymbol{E} \cdot \boldsymbol{j}
\label{Eq.6}
\end{eqnarray}
where $\boldsymbol{j}=\sum_{\alpha} \boldsymbol{j}_\alpha$ is the total electric current density, and
$\boldsymbol{j}_\alpha=n_\alpha q_\alpha \boldsymbol{u}_\alpha$ is the electric current density of species $\alpha$.
Under certain boundary conditions,
e.g., periodic,
integrating Eqs. {\ref{Eq.4}}, {\ref{Eq.5}}, and {\ref{Eq.6}} over the whole volume,
we can have
\begin{eqnarray}
\partial_t \langle E^{f}_\alpha \rangle &=& \langle \left( \boldsymbol{P}_\alpha \cdot \nabla \right) \cdot \boldsymbol{u}_\alpha \rangle+ \langle n_\alpha q_\alpha \boldsymbol{E} \cdot \boldsymbol{u}_\alpha \rangle, \label{Eq.energy-conversion1}\\
\partial_t \langle E^{th}_\alpha \rangle &=& -\langle \left( \boldsymbol{P}_\alpha \cdot \nabla \right) \cdot \boldsymbol{u}_\alpha \rangle, \label{Eq.energy-conversion2}\\
\partial_t \langle E^{m} \rangle &=& - \langle \boldsymbol{E} \cdot \boldsymbol{j} \rangle. \label{Eq.energy-conversion3}
\end{eqnarray}
where $\langle \cdots \rangle$ denotes a space average
over the entire volume.

\begin{figure}[!htpb]
\centering
\begin{picture}(12,7.75)
\put(3.5,0){\framebox(5,2){electromagnetic energy $\langle E^{m} \rangle$}}
\put(-1.5,4){\framebox(5.5,2){thermal (random) energy $\langle E^{th}_\alpha \rangle$}}
\put(8,4){\framebox(4,2){fluid flow energy $\langle E^{f}_\alpha \rangle$}}
\put(4,4.8){\vector(1,0){4}}\put(8,5.2){\vector(-1,0){4}}
\put(4.7,5.4){$-\langle \left( \boldsymbol{P}_\alpha \cdot \nabla \right) \cdot \boldsymbol{u}_\alpha \rangle$}
\put(9.5,4){\vector(-2,-1){4}}\put(6.5,2){\vector(2,1){4}}
\put(8.6,2.8){$\langle \boldsymbol{j}_\alpha \cdot \boldsymbol{E} \rangle$}
\end{picture}
\caption{Illustration of the available
routes for global energy conversion.
The point-wise values of
$-\left( \boldsymbol{P}_\alpha \cdot \nabla \right) \cdot \boldsymbol{u}_\alpha$
and $\boldsymbol{j}_\alpha \cdot \boldsymbol{E}$ are not sign-definite.
Therefore, there are two possible directions of energy conversion.
$\langle \cdots \rangle$ denotes the space average over the entire volume.}
\label{Fig.energy conversion}
\end{figure}
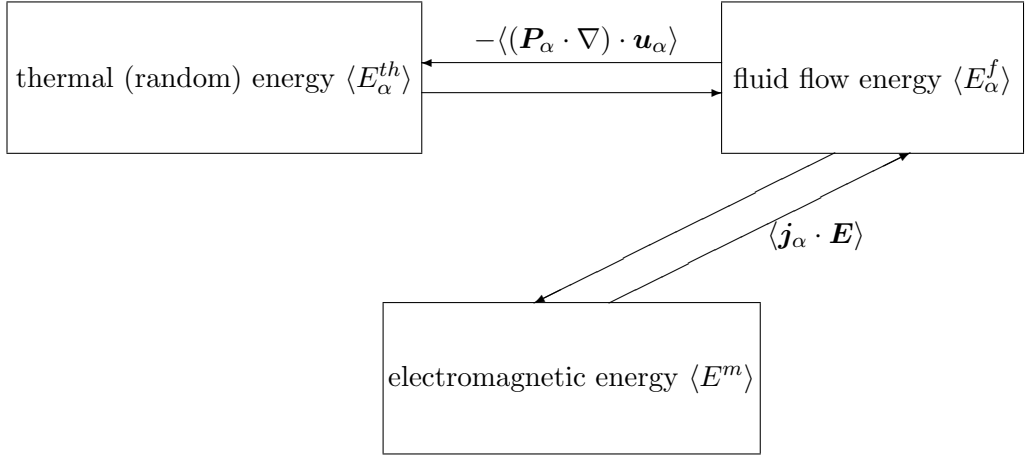

Fig. {\ref{Fig.energy conversion}} illustrates
the energy conversions as suggested by Eqs. {\ref{Eq.energy-conversion1}},
{\ref{Eq.energy-conversion2}} and {\ref{Eq.energy-conversion3}}.
One can see that for the collisionless case derived from the Vlasov equation,
the pressure work,
$-\langle \left( \boldsymbol{P}_\alpha \cdot \nabla \right) \cdot \boldsymbol{u}_\alpha \rangle$,
is the only term converting fluid flow energy into thermal (random) energy,
while the term,
$ \langle \boldsymbol{j}_\alpha \cdot \boldsymbol{E}\rangle$,
represents the conversion between fluid flow and electromagnetic energies.
The pressure work
seems to be a more straightforward measure
of heating rate when compared with
electromagnetic work {\citep{WanEA15,WanEA16}}.
At present, we cannot rule out the possibly strong correlation
between the work done by pressure and
work done by the electric field.
For example, for
a generalized
Ohm's law, or the electron momentum equation, in the limit of
massless electrons in collisionless plasma,
we find that
$\langle \boldsymbol{j}_e \cdot \boldsymbol{E} \rangle = -\langle \left( \boldsymbol{P}_e \cdot \nabla \right) \cdot \boldsymbol{u}_e \rangle$.

Notwithstanding that the pressure work
is a general property
in various fluid systems,
seldom have studies investigated
the role of pressure in modifying
the thermal (random) energy
in a turbulent kinetic plasma (however, see, e.g.,
\citet{BirnHesse01,BirnEA06,BirnHesse10}).
In the realm of observations
this is more or less due to the
intractability of calculating velocity gradient
from single spacecraft datasets and until recently,
a lack of high cadence determination of the pressure
tensor. These complication have
led most observational studies of solar wind turbulence to
rely on high cadence magnetic field data, which is generally much more
accessible. Even in simulation studies,
accurate determination of the pressure
tensor is challenging, requiring either large
numbers of particles in PIC codes,
or the use of computationally
demanding Eulerian Vlasov simulations.
Here we will use numerical simulations to explore the
role of pressure tensor in heating of kinetic plasma in detail.

\section{Simulation details}
The fully kinetic particle-in-cell (PIC) simulation employed here
spans from macroscopic fluid scales to kinetic scales.
It is expedient for studies of energy transfer and dissipation.
The simulation was performed using
P3D code {\citep{Zeiler02}} in 2.5D geometry
(three components of dependent field vectors and
a two-dimensional spatial grid).
Number density is normalized to the reference
number density $n_r$ (=1 in this simulation),
mass to proton mass $m_i$ (=1 in this simulation), and
magnetic field to $B_r$ (=1 in this run).
Length is normalized to the ion inertial length $d_i$,
time to the ion cyclotron time $\Omega_i^{-1}$, and
velocity to the reference Alfv{\'e}n speed
$v_{Ar}=B_r/\left(4\pi m_i n_r\right)^{1/2}$.
Parameters of the PIC simulation
are listed in Table {\ref{Table.Params}}.
It is conducted in periodic boundary conditions in both directions.
The initial density, temperature and
out-of-plane magnetic field $B_0$
are uniform.
The initial $v$ and $b$ fluctuations are transverse to $B_0$
(``Alfv\'en mode'') and
excited for prescribed wavenumbers
with specified spectra and cross helicity.
More details about the simulation can be found in Ref. {\citet{WuEA13-vKH}}.
\begin{table}[!htpb]
\small
\setlength{\belowcaptionskip}{5pt}
\centering
\caption{Simulation parameters: box size $L$, grid points $N^2$, mass ratio $m_i/m_e$, proton beta $\beta_i$,
electron beta $\beta_e$, out of plane uniform magnetic field $B_0$, the number of particles of each species
per grid $ppg$
and correlation scale $\lambda_c$. }
\begin{tabular*}{\textwidth}{@{\extracolsep{\fill}} ccccccccc}
\hline
Code & $L$ & $N^2$ & $m_i/m_e$ & $\beta_i$ & $\beta_e$ & $B_0$ & $ppg$ & $\lambda_c$\\
\hline

P3D{\citep{Zeiler02}} & $102.4d_i$ & $8192^2$ & 25 & 0.1 & 0.1 & 5.0 & 300 & $16.8d_i$\\

\hline
\end{tabular*}
\label{Table.Params}
\end{table}

\begin{figure}[!htpb]
\centering
\includegraphics[width=0.45\textwidth]{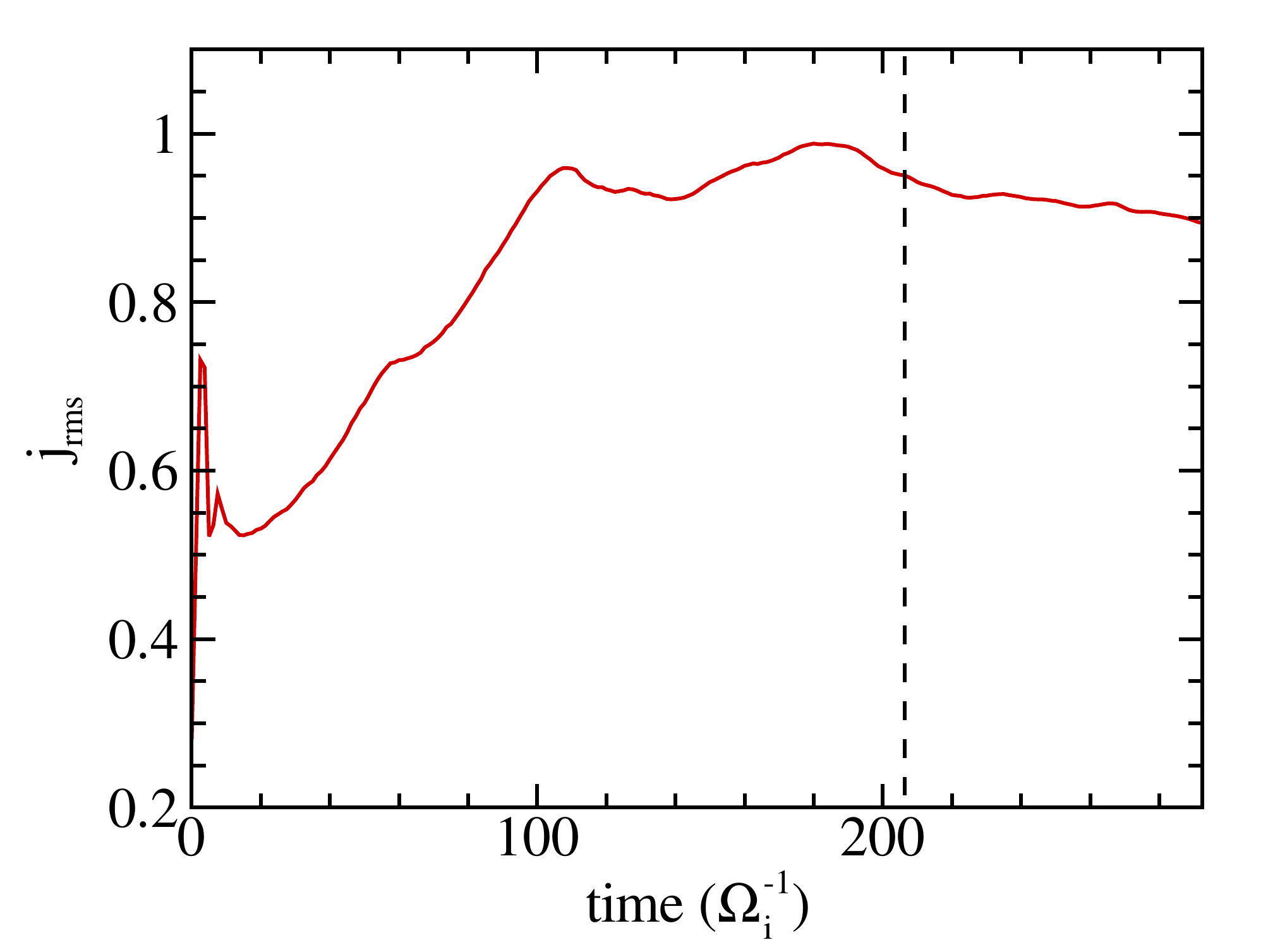}
\includegraphics[width=0.45\textwidth]{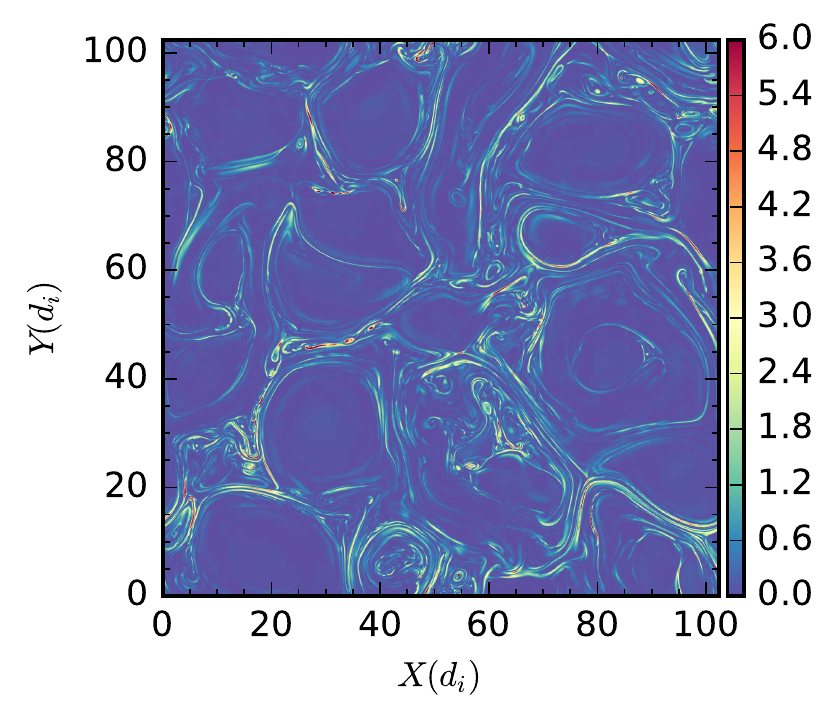}
\caption{(a) Time history of root mean square(r.m.s.)
electric current density $j_{rms}$.
The time snapshot ($t\Omega_{i}=206.25$)
near the maximum r.m.s. total electric current density,
which is indicated by the dashed line, is analyzed.
(b) Contour of the normalized
second ``invariant'' of current density
$Q_j={\frac{1}{4}}\boldsymbol{j}^2/ \langle \boldsymbol{j}^2 \rangle$.}
\label{Fig.time-evolution-current}
\end{figure}

We analyze statistics near the time
of maximum root mean square(r.m.s.) electric current density
(i.e., $t\Omega_{i}=206.25$),
shown by the dashed lines in Fig. {\ref{Fig.time-evolution-current}}(a).
By this time the turbulence has fully developed and
it has generated a complex pattern,
characterized by sheet-like current structures spanning
a range of scales as seen in Fig. {\ref{Fig.time-evolution-current}}(b).
In order to remove particle noise at grid scales, we apply
a low-pass Fourier filter
to the field data for $k\lambda_d>1$
(Debye length $\lambda_d$)
prior to statistical analyses.
Demonstration of the efficacy of this filtering
technique will be documented elsewhere \cite{HaggertyEA17}.

\section{Role of pressure tensor}
\label{Sec.RolePressureTensor}
A somewhat surprising result \cite{YangEA-PRL-17}
further documented here is the
correlation of pressure effects and
dynamically appearing coherent structures associated with
intermittency.
In order to clarify the correlation between
pressure tensor and intermittent structures,
we split the pressure work,
$-\left( \boldsymbol{P} \cdot \nabla \right) \cdot \boldsymbol{u}$,
into two parts as follows.
We suppress the subscript $\alpha$ for simplicity.

Decomposing the pressure tensor, $\boldsymbol{P}=\left(P_{ij}\right)$, gives
\begin{equation}
P_{ij}=p\delta_{ij}+\Pi_{ij},
\label{Eq.splitP}
\end{equation}
where $p={\frac{1}{3}}P_{ii}$ is the scalar pressure;
a sum on repeated indexes is implied;
$\delta_{ij}$ is the Kronecker delta;
$\Pi_{ij}$ is the remaining part with the scalar pressure
subtracted from the pressure tensor,
that is, the deviatoric pressure tensor.

The intrinsic decomposition of $\nabla \boldsymbol{u}$
into symmetric and anti-symmetric parts gives
\begin{eqnarray}
\partial_i u_j &=& S_{ij}+\Omega_{ij}, \nonumber\\
&=& {\frac{1}{3}}\theta \delta_{ij} + D_{ij}+ \Omega_{ij},
\label{Eq.splitV}
\end{eqnarray}
where $S_{ij}={\frac{1}{2}}\left(\partial_i u_j + \partial_j u_i\right)$
and $\Omega_{ij}={\frac{1}{2}}\left(\partial_i u_j - \partial_j u_i\right)$, with $\epsilon_{ijk}\Omega_{jk}=\omega_i$, are the strain-rate and rotation-rate tensors, respectively; $\omega_{i}$ is the vorticity; $\theta=S_{ii}$ is the dilatation; $D_{ij}=S_{ij}-{\frac{1}{3}}\theta \delta_{ij}$ is the traceless strain-rate tensor.

Using Eqs. {\ref{Eq.splitP}} and {\ref{Eq.splitV}}, we can obtain
\begin{equation}
-\left(\boldsymbol{P} \cdot \nabla \right) \cdot \boldsymbol{u} = -p\theta-\Pi_{ij}D_{ij}.
\end{equation}
One can see that $-p\theta$, accounting for compressive effects,
is of the same form as the pressure dilatation in compressible MHD turbulence.
The new term in kinetic plasma in comparison with MHD is
$-\Pi_{ij}D_{ij}$, the double contraction of
deviatoric pressure tensor and traceless strain-rate tensor.
This term which will be called ``{\it Pi-D}'' hereafter
is a salient feature of the present paper.

At this point we may recall that
in the continuum formulation
leading to the Navier-Stokes equations,
with strong collisions,
$-\Pi_{ij}$ actually is present,
but rarely written in this way.
Instead, in a Chapman-Enskog development,
this term is equated with a viscous stress,
which can then be expressed in terms of
velocity gradient {\citep{Vincenti65,Braginskii65}}.
In Sec. {\ref{Sec:PD-coherent-structures}},
we will check whether the
``{\it Pi-D}'' term shares some properties
with viscous dissipation.

Table {\ref{Table.PD}} contains analysis based on the
numerical simulation
that compares $\langle -p\theta \rangle$ and
$\langle -\Pi_{ij}D_{ij} \rangle$
and the electromagnetic work.
One can see that
the pressure dilatation is smaller
as expected based on the weak compressibility in the run,
where $\delta \rho'_\alpha/\langle\rho_\alpha\rangle=
\sqrt{\langle(\rho_\alpha-\langle\rho_\alpha\rangle)^2\rangle}/\langle\rho_\alpha\rangle \simeq 0.12$.
The $\langle -\Pi_{ij}D_{ij} \rangle$ terms are comparable to $\langle \boldsymbol{E} \cdot \boldsymbol{j}_\alpha \rangle$.
Time average over about an electron gyroperiod is used in
computing $\langle \boldsymbol{E} \cdot \boldsymbol{j}_\alpha \rangle$
to eliminate very high frequency oscillations.

\begin{table}[!htpb]
\small
\setlength{\belowcaptionskip}{5pt}
\centering
\caption{Strength of global conversion of the fluid flow, thermal (random) and electromagnetic energies.
All quantities are listed in the unit $v_{Ar}^3 d_i^{-1}$.}
\begin{tabular*}{\textwidth}{@{\extracolsep{\fill}}|c|ccc|}
\hline
Species & $\langle -p\theta \rangle$ & $\langle -\Pi_{ij}D_{ij} \rangle$ & $\langle \boldsymbol{E} \cdot \boldsymbol{j}_\alpha \rangle$\\
\hline

Electron & 0.0018  & 0.0045 & 0.0052\\

\hline

Ion & 0.00075  & 0.0016 & 0.0016\\

\hline

\end{tabular*}
\label{Table.PD}
\end{table}

\begin{figure}[!htpb]
\centering
\includegraphics[width=0.9\textwidth]{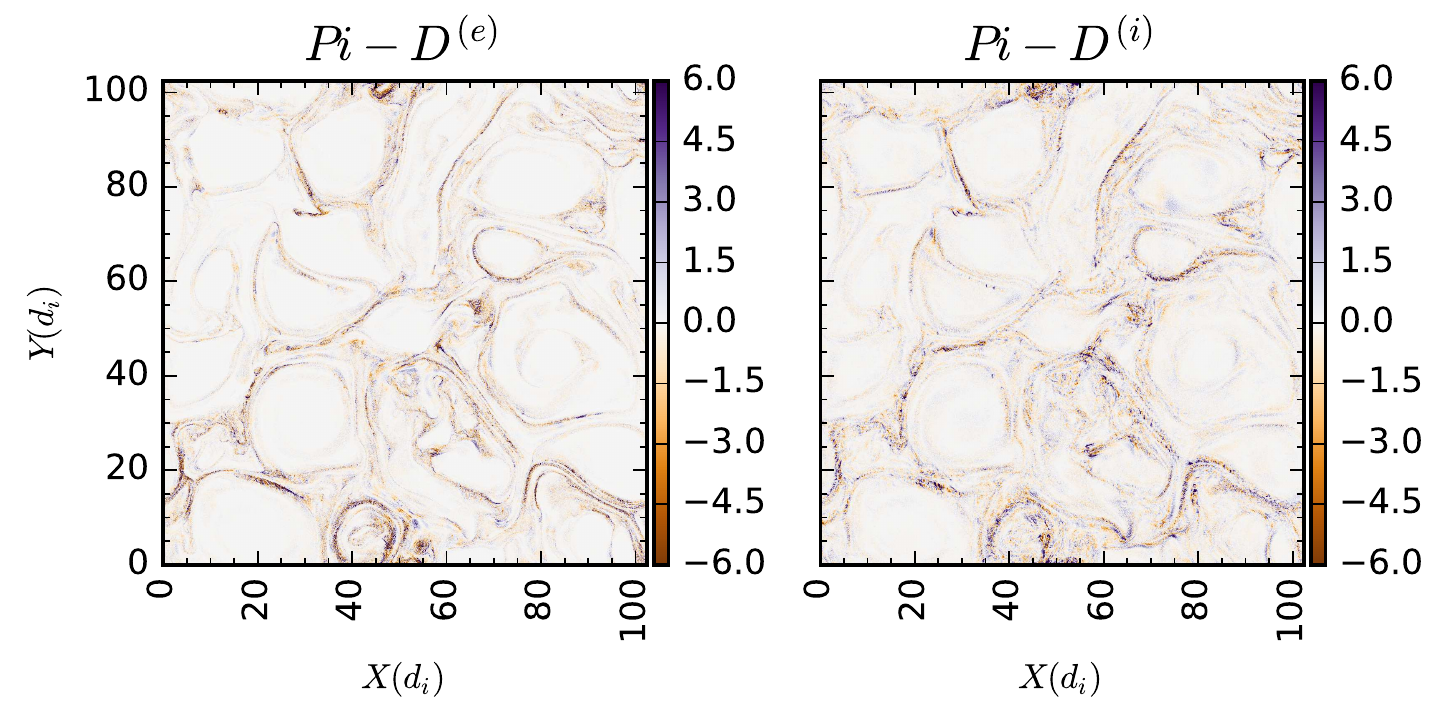}
\includegraphics[width=0.45\textwidth]{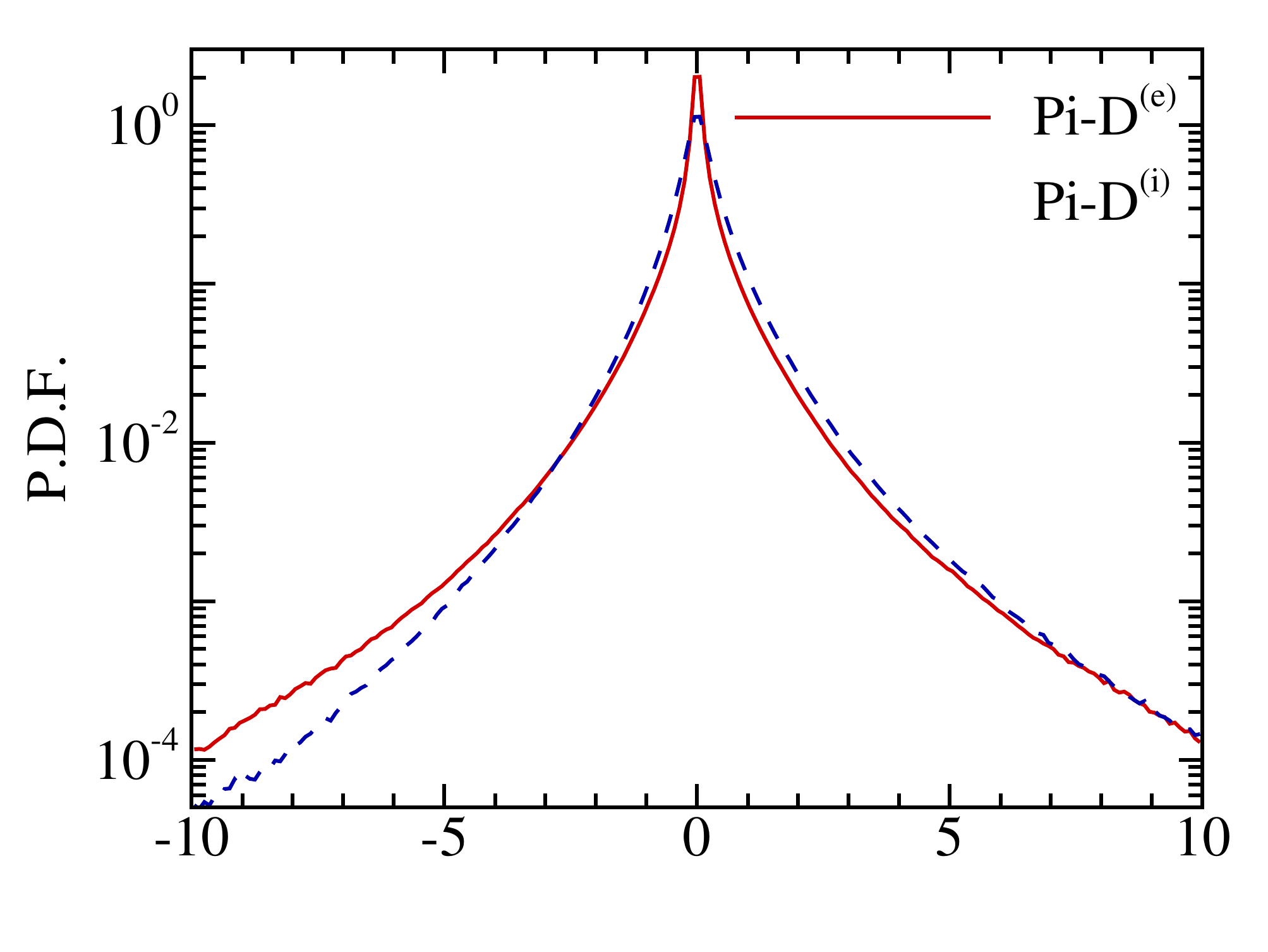}
\caption{(a) Contours of $-\Pi_{ij}D_{ij}$ for electrons (left) and ions (right)
(reproduced from Paper I for completeness).
Both quantities are normalized to their respective root mean square values.
Both are organized into sheet-like structures.
They are not sign-definite, thus the directions of energy conversion vary.
(b) PDFs of normalized $-\Pi_{ij}D_{ij}$ of both electrons (red solid) and ions (blue dashed).
They slightly tilt towards positive values, indicating that
the fluid flow energy is converted into thermal (random)
energy globally.}
\label{Fig.ContPD}
\end{figure}

To clarify the effect of the pressure tensor,
we plot in Fig. {\ref{Fig.ContPD}}(a) the contours of ``{\it Pi-D}'' for
both electrons and ions.
The ``{\it Pi-D}'' term contributes substantially
to the energy exchange between the fluid flow
and the random kinetic energy. These contributions are
concentrated locally in space.
The intensity and signs of ``{\it Pi-D}'' vary,
so also do the amount and direction
of the energy conversion.
It is intermittent,
while its net effect over the whole domain,
i.e., its global average, is relatively small,
which is also verified through the PDF plot
in Fig. {\ref{Fig.ContPD}}(b).
The long-tailed curves there
slightly tilt towards positive values,
i.e., the fluid flow energy is converted into
thermal (random) energy globally.

\begin{figure}[!htpb]
\centering
\includegraphics[width=0.9\textwidth]{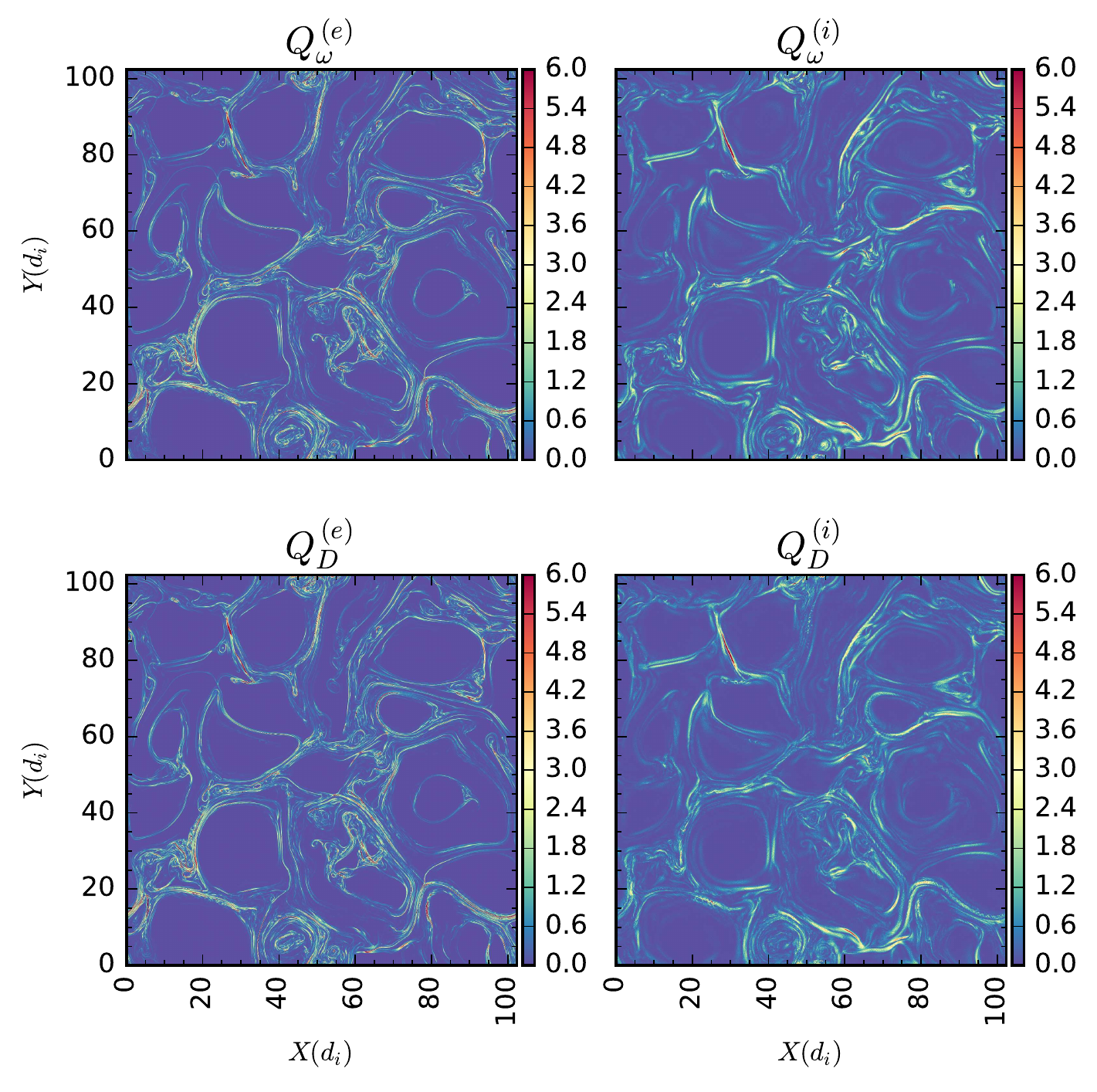}
\caption{Contours of
the normalized second invariant of the rotation-rate tensor, (top row),
$Q_\omega={\frac{1}{4}}\boldsymbol{\omega}^2/ \langle \boldsymbol{\omega}^2 \rangle$,
and the normalized second invariant of the traceless strain-rate tensor (bottom row),
$Q_D={\frac{1}{2}}D_{ij}D{ij}/ \langle 2D_{ij}D_{ij} \rangle$.
Left panels: electrons (reproduced from Paper I for completeness); and right panels: ions.}
\label{Fig.ContQWQD}
\end{figure}

\subsection{Energy conversion related to coherent structures}
\label{Sec:PD-coherent-structures}
Various studies based on
numerical simulations
{\citep{ParasharEA11,Servidio12,Greco12,WanEA12-prl,Karimabadi13,TenBarge13,WanEA15}}
and solar wind data
{\citep{Sundkvist07,Osman11,Osman12a,Osman12b,WuEA13-SW}}
support the idea that
enhanced kinetic activity, such as
temperature anisotropy, heating, particle acceleration, and
departures from Maxwellian velocity distributions in general,
all of which commonly observed in astrophysical and laboratory plasmas,
are strongly inhomogeneous.
These effects
are associated typically
with coherent structures such as
magnetic structures.
Indeed,
intense kinetic activity is often found
in the general proximity to strong gradients,
including not only magnetic,
but also
strong density and velocity gradients
{\citep{Huba96, Markovskii06, Vasquez12,
Servidio15, Franci16, DelSartoPegoraro16, ParasharMatthaeus16}}.
In particular, Refs. {\citep{ServidioEA14-Vlasov,Franci16, ParasharMatthaeus16}}
find that heating is correlated with both
vorticity and current density, but more strongly
with vorticity.

Here we are interested in looking at velocity gradients
and their interaction with the pressure tensor,
in view of the important role
of these quantities in energy conversion
in kinetic plasma,  as seen in Eq. \ref{Eq.5}.
We base our diagnostics on
the geometric invariants of the relevant
second-order tensorial quantities,
an approach extensively employed
in hydrodynamics
to describe flow patterns \citep{Chong90, Soria94, Chong98, Ooi99}.
Based on the decomposition in Eq. {\ref{Eq.splitV}},
the normalized second invariants
of
the traceless strain-rate matrix $\left(D_{ij}\right)$
and the rotation-rate matrix $\left(\Omega_{ij}\right)$
are $Q_D={\frac{1}{2}}D_{ij}D_{ij}/ \langle 2D_{ij}D_{ij} \rangle$
and $Q_\omega={\frac{1}{4}}\boldsymbol{\omega}^2/ \langle \boldsymbol{\omega}^2 \rangle$,
respectively.
We can also define a similar quantity
for the electric current density, say,
$Q_j={\frac{1}{4}}\boldsymbol{j}^2/ \langle \boldsymbol{j}^2 \rangle$.

Figs. {\ref{Fig.ContQWQD}} and {\ref{Fig.time-evolution-current}}(b) show
contours of $Q_\omega$, $Q_D$ and $Q_j$,
which are found to be non-uniformly distributed in space.
Moreover, comparison of Figs. {\ref{Fig.ContQWQD}}, {\ref{Fig.time-evolution-current}}(b)
with Fig. {\ref{Fig.ContPD}}(a) reveals greatly similar patterns
of {\it Pi-D}, $Q_\omega$, $Q_D$ and $Q_j$,
as also shown in {\citep{ParasharMatthaeus16}}. This indicates,
as also described in Paper I, that
these intermittent structures might be sites of enhanced
energy conversion,
which is consistent with various recent results
{\citep{Dmitruk04,Retino07,Sundkvist07,ParasharEA11,TenBarge13,Perri12,
ServidioEA14-Vlasov,Servidio15}}.

\begin{figure}[!htpb]
\centering
\includegraphics[width=0.9\textwidth]{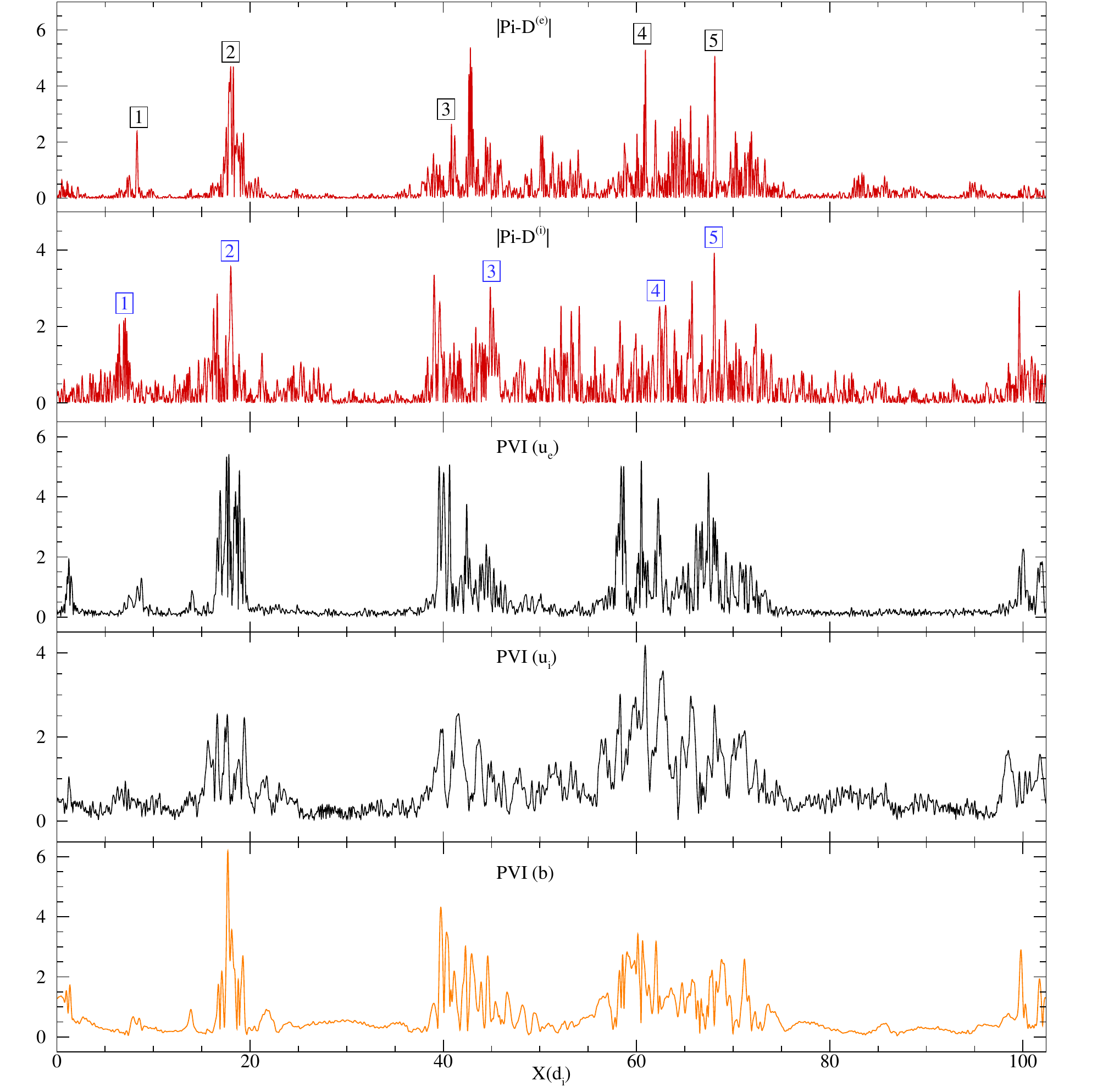}
\caption{Series of values on a cut ($Y \simeq 35 d_i$) in $X$ direction.
(Top to bottom) absolute values of normalized ``{\it Pi-D}'' for electrons;
absolute values of normalized ``{\it Pi-D}'' for ions;
PVI values for electron bulk velocity ($\boldsymbol{u}_e$);
PVI values for ion bulk velocity ($\boldsymbol{u}_i$);
PVI values for magnetic field ($\boldsymbol{b}$).
Events labeled by sequential numbers
indicate highly intermittent regions that are enlarged below.}
\label{Fig.CutPD}
\end{figure}

\begin{figure}[!htpb]
\centering
\includegraphics[width=0.9\textwidth]{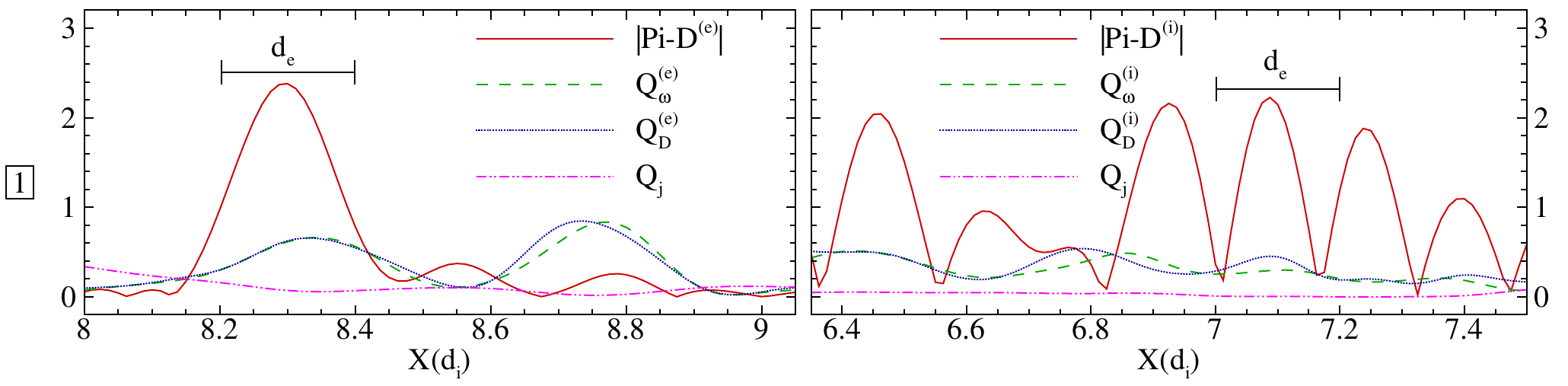}
\includegraphics[width=0.9\textwidth]{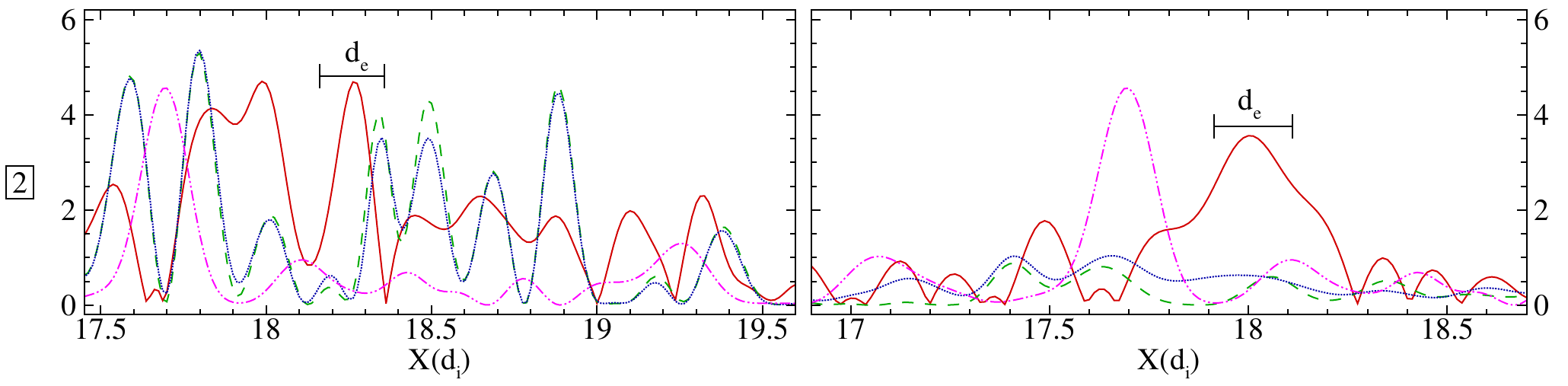}
\includegraphics[width=0.9\textwidth]{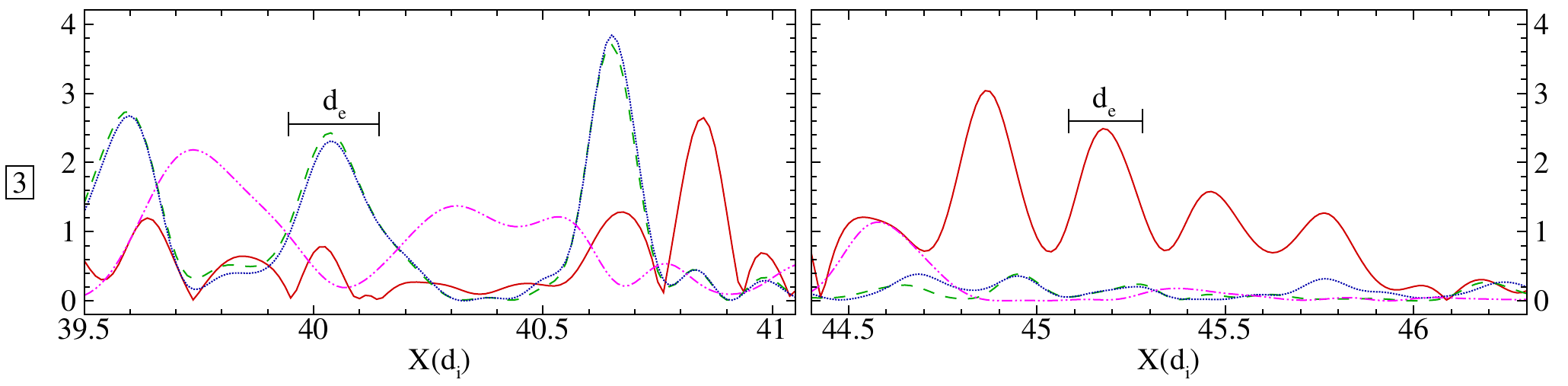}
\includegraphics[width=0.9\textwidth]{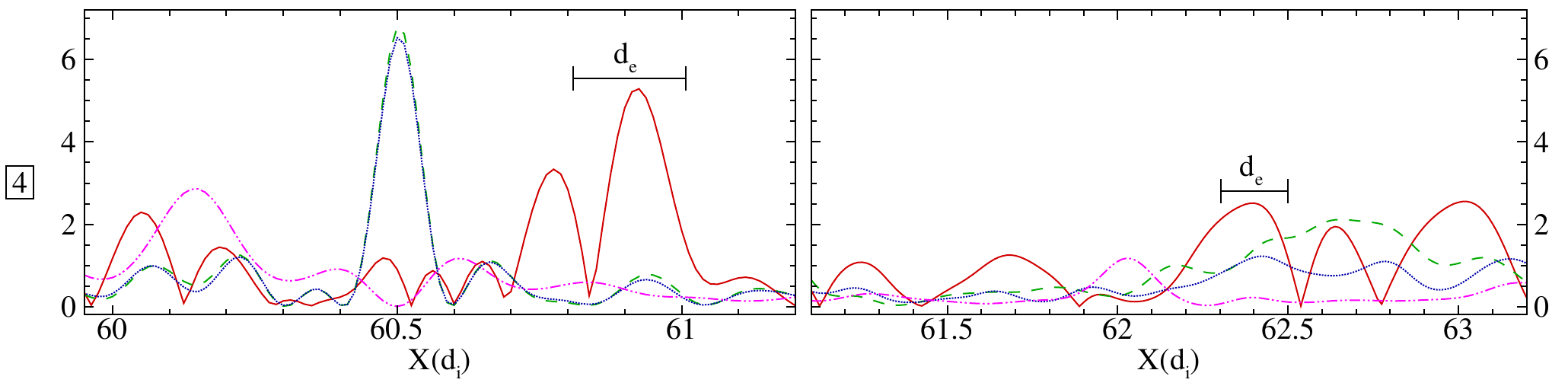}
\includegraphics[width=0.9\textwidth]{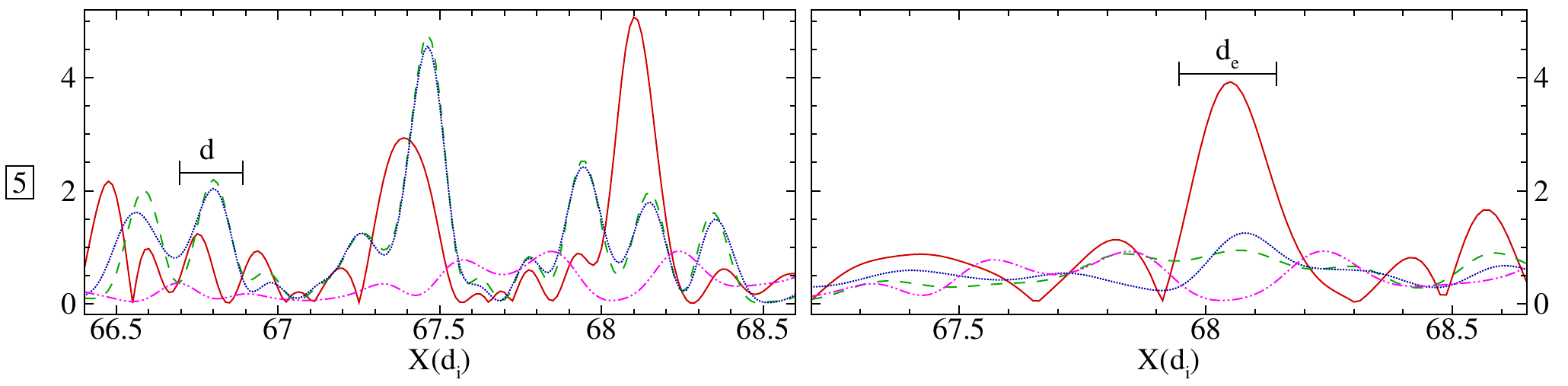}
\caption{Quantities, $\left|-\Pi_{ij}D_{ij}/\left(-\Pi_{ij}D_{ij}\right)_{rms} \right|$,
$Q_\omega$ and $Q_D$ for both (left) electrons and (right) ions,
and $Q_j$ from total current density. Top to bottom: regions labeled
with 1 to 5 in Fig. {\ref{Fig.CutPD}}. The black horizontal line
shows the electron inertial scale $d_e$.}
\label{Fig.ZoomCut}
\end{figure}

We highlight the possible correlations of
{\it Pi-D}, $Q_\omega$, $Q_D$ and $Q_j$
by plotting them on cuts along the $X$ direction.
A sample of the absolute values of ``{\it Pi-D}'' terms
along the $X$ direction with $Y \simeq 35d_i$ is shown
in Fig. {\ref{Fig.CutPD}}.
One sees the spatial distributions of ``{\it Pi-D}'' terms
are evidently bursty, suggesting a connection to
the spatial intermittency of the turbulence.
A useful intermittency measure is given by
the partial variance of increments (PVI) \citep{Greco12},
${\rm{PVI}} (\boldsymbol{f})= {\frac{\left|\Delta \boldsymbol{f}\right|}
{\sqrt{\langle\left|\Delta \boldsymbol{f}\right|^2\rangle}}}$, where
$\Delta \boldsymbol{f} = \boldsymbol{f}(s+ \Delta s) - \boldsymbol{f}(s)$.
Here we choose a small scale lag, $\Delta s \simeq 0.2d_i = d_e$.
The PVI series of bulk velocities for electrons ($\boldsymbol{u}_e$)
and ions ($\boldsymbol{u}_i$) and magnetic field ($\boldsymbol{b}$)
in Fig. {\ref{Fig.CutPD}}
behave quite similar to ``{\it Pi-D}'',
with peak values in the vicinity of
high ``{\it Pi-D}''.
Events with high ``{\it Pi-D}'' on the series
are selected by labeling with sequential numbers in the figure,
and also displayed along with $Q_\omega$, $Q_D$ and $Q_j$ in Fig. {\ref{Fig.ZoomCut}}.
We might expect that intense kinetic activities
are associated with current sheets, in particular,
with high values of current density.
In Fig. {\ref{Fig.ZoomCut}}, however, the energy conversion through ``{\it Pi-D}''
is more correlated with $Q_\omega$ and $Q_D$ in comparison with $Q_j$.
This can be seen from the fluctuations for electrons
in the left panels of Fig. {\ref{Fig.ZoomCut}}, where local maxima of
the absolute ``{\it Pi-D}'' terms and $Q_D$ (or $Q_\omega$)
are close to each other, indicating a well-established association
between the symmetric and antisymmetric parts of velocity gradients
and the energy conversion through the deviatoric pressure work.
This result further strengthens the idea of
{\citep{Servidio15, DelSartoPegoraro16, Franci16, ParasharMatthaeus16}}
that energization occurs near to, but not centered on, current sheets,
and that regions with large
velocity gradients are prime
locations for energy exchange between fields and particles.

The possible correlation between {\it Pi-D}
and $Q_D$ implies more.
As is well known,
the viscous dissipation in HD and MHD turbulence is
proportional to the mean square gradient of velocity,
thus
$Q_D$ is a surrogate of the viscous dissipation
in collisional fluid.
This result suggests a possible resemblance
between collisionless and collisional (viscous) dissipation
functions.
Note that the deviatoric
pressure work, i.e., ``{\it Pi-D}'',
differs from an irreversible dissipation
mechanism, since the pointwise {\it Pi-D},
as shown in Fig. {\ref{Fig.ContPD}},
is not positive-definite.
A pointwise negative value of {\it Pi-D}
means that thermal (random) energy is
converted into flow kinetic energy;
conversely positive values of the pointwise
{\it Pi-D} imply
a positive time rate of change of thermal (random)
energy.
In spite of this fundamental
difference,
the global energy conversion in this case is
of approximately fluid-like form, which
lends credence again to the idea suggested by
Vasquez et al. {\citep{Vasquez12}},
the kinetic heating of protons
might be a ``viscous like'' process instead of a magnetic process.
A possible clue comes from recent work
{\citep{DelSartoPegoraro16}} that
describes in detail
how the eigenvalues of the velocity gradient tensor
contribute to the time derivative of the
pressure tensor
(see, e.g., Eq.(15) of Ref. {\citep{DelSartoPegoraro16}}).
In the light of this, one can expect ``{\it Pi-D}''
to be correlated with $Q_D$ more strongly
in comparison with $Q_\omega$,
as is the case for ions in the right panels of Fig. {\ref{Fig.ZoomCut}}.

The net effect of the deviatoric pressure work in our case
is found to increase internal energy as was done by dissipation
in collisional fluids.
This provides a possible pathway
when collisional relaxation is either absent or weak
in kinetic plasma to understand
how  part of the fluid flow energy is converted
into thermal (random) energy.
To arrive at this perspective,
there is no need
to specify a particular mechanism,
such as reconnection heating, Landau damping,
cyclotron damping, or stochastic heating. Since the above commentary is
based on the Vlasov equation alone,
any process that contributes to the net ``{\it Pi-D}''
interaction is contributing to thermal energy increase.
We have also not attempted to address the question of
what physical processes make such energy exchanges irreversible.
Hence we cannot comment on
how much of the energy converted into random motions
can settle down to heat permanently.
With these caveats in mind,
we defer these questions to future study.

\subsection{Joint PDFs of coherent structures}
\begin{figure}[!htpb]
\centering
\includegraphics[width=0.45\textwidth]{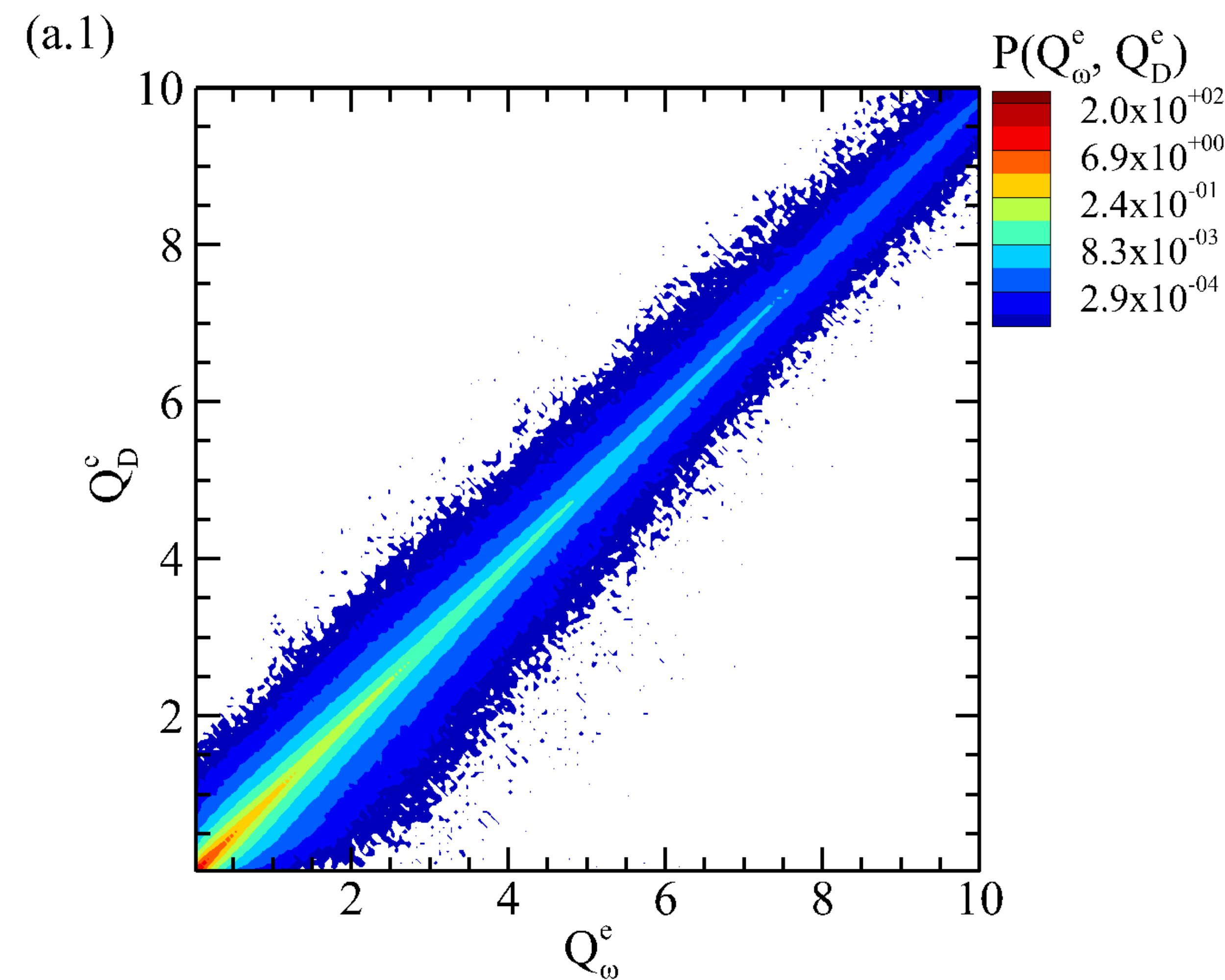}
\includegraphics[width=0.45\textwidth]{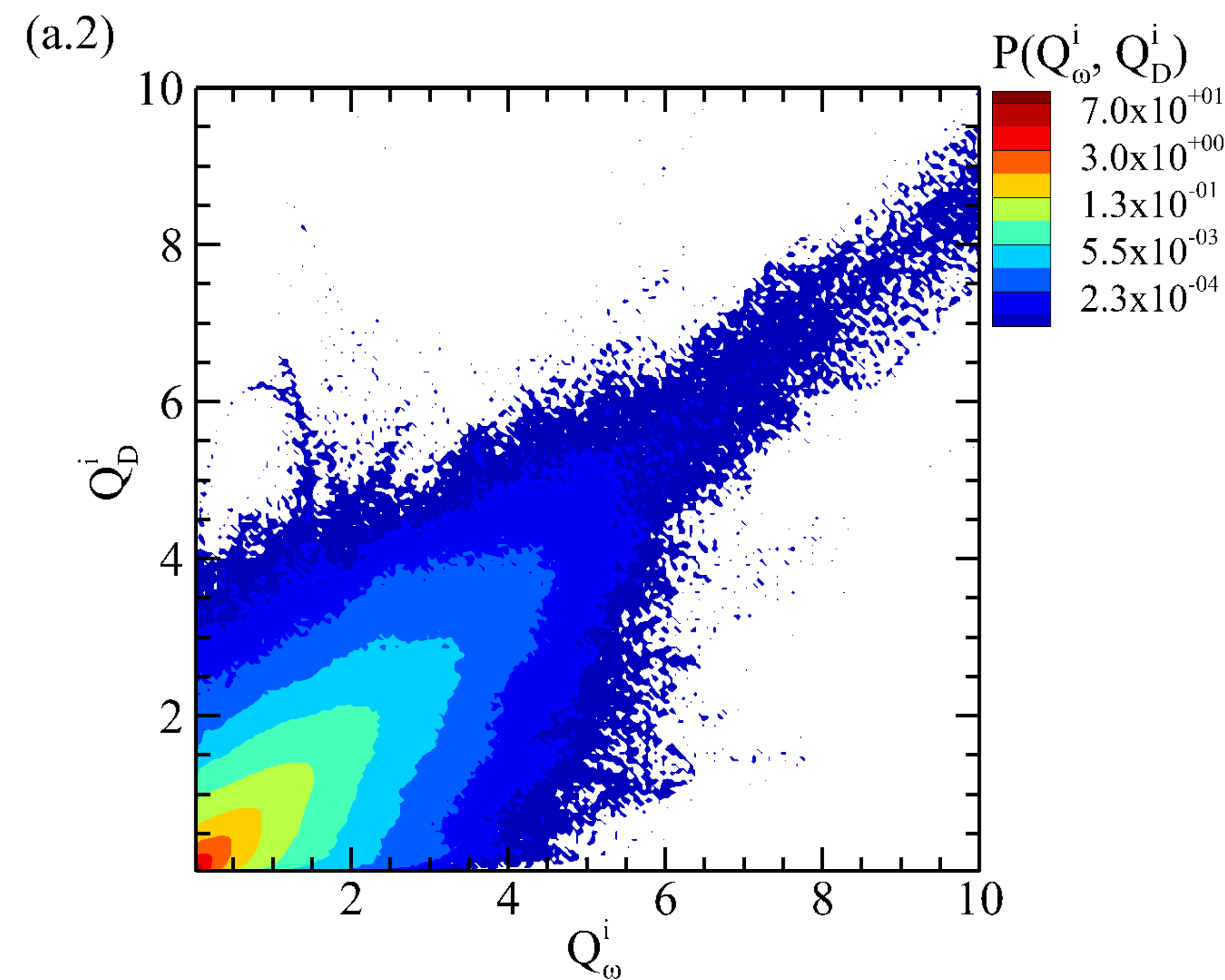}
\includegraphics[width=0.45\textwidth]{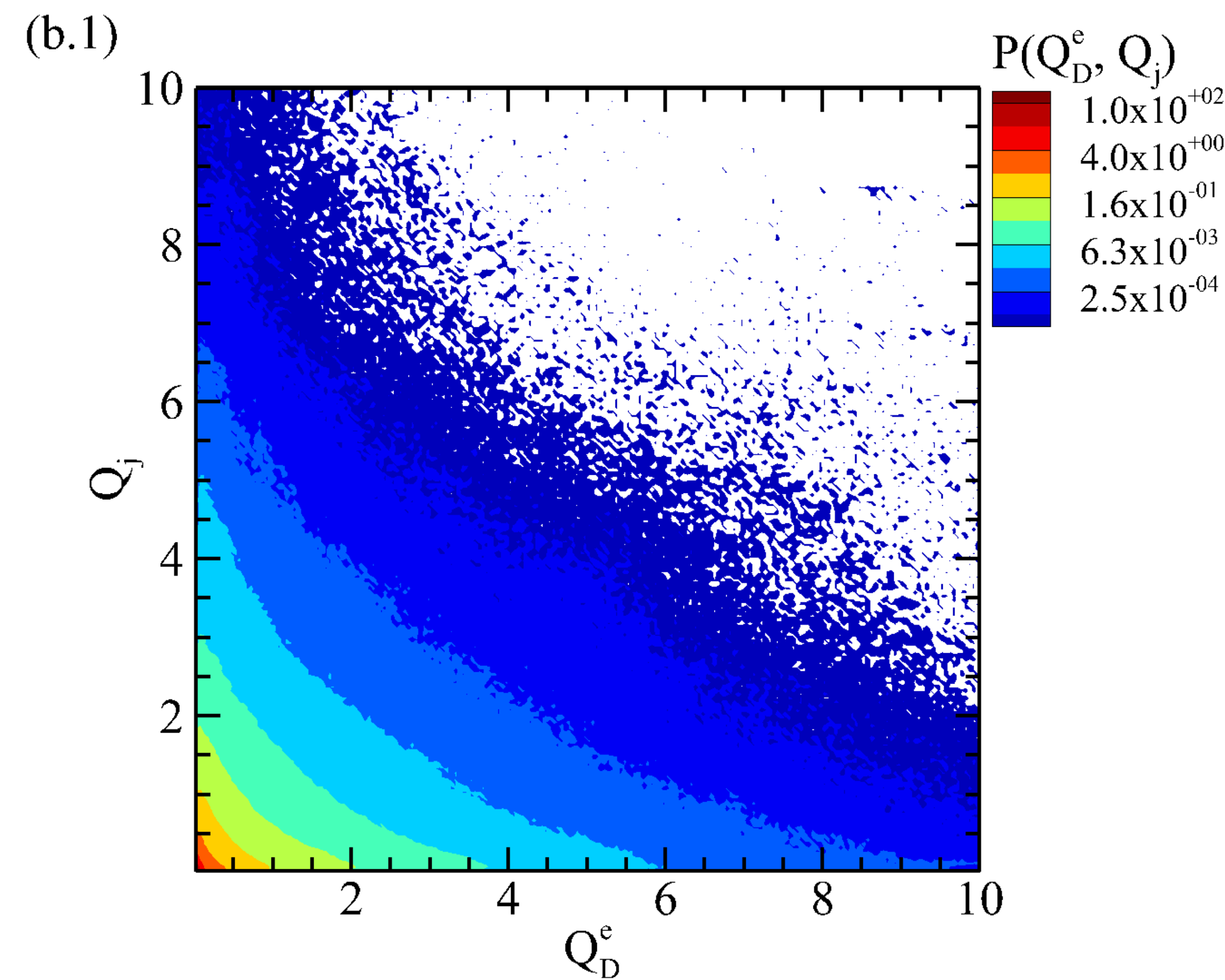}
\includegraphics[width=0.45\textwidth]{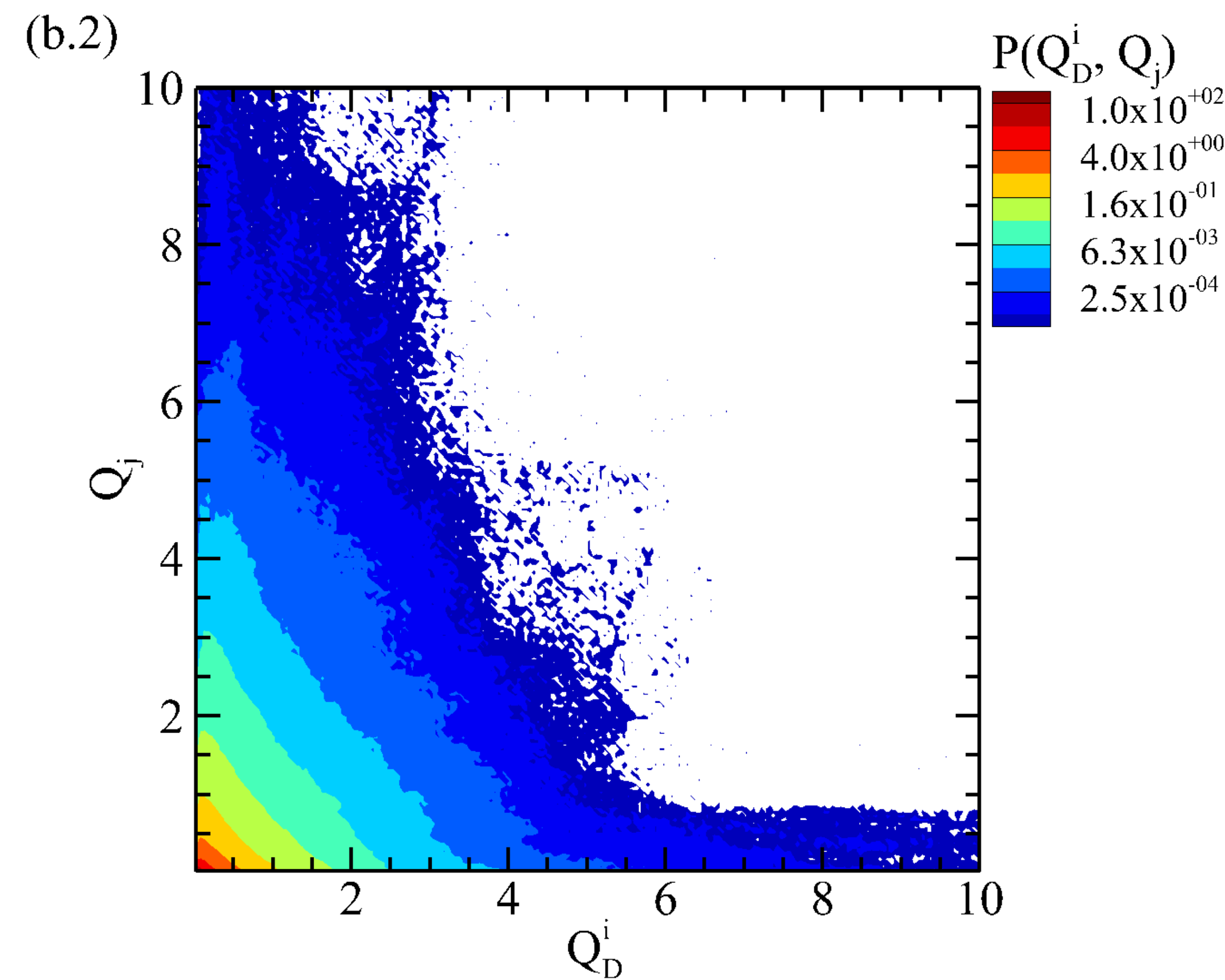}
\includegraphics[width=0.45\textwidth]{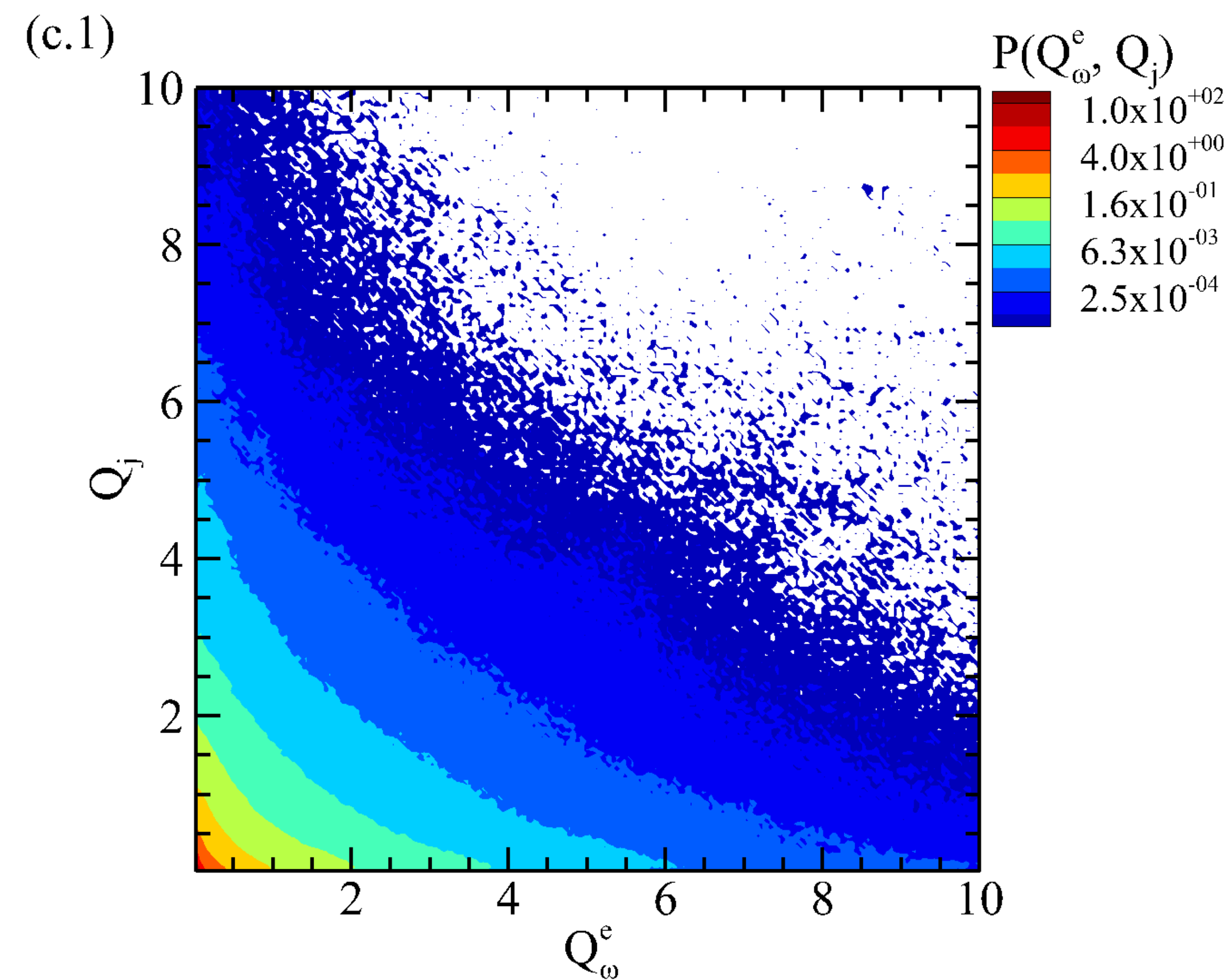}
\includegraphics[width=0.45\textwidth]{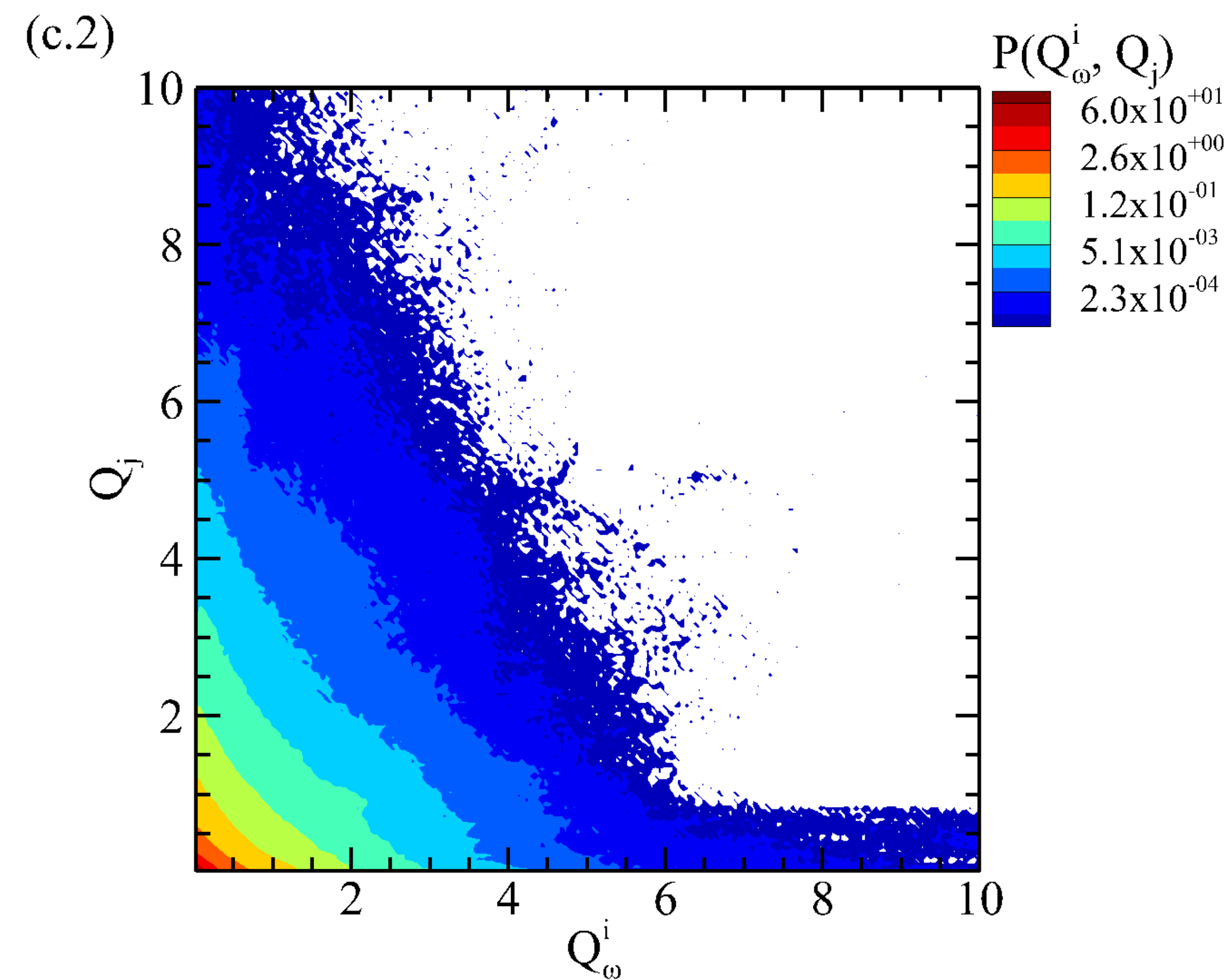}
\caption{Joint PDFs $P(x,y)$ of
the normalized second invariants of rotation-rate, traceless strain-rate tensors,
and current density,
i.e., $Q_\omega={\frac{1}{4}}\boldsymbol{\omega}^2/ \langle \boldsymbol{\omega}^2 \rangle$,
$Q_D={\frac{1}{2}}D_{ij}D{ij}/ \langle 2D_{ij}D_{ij} \rangle$,
and $Q_j={\frac{1}{4}}\boldsymbol{j}^2/ \langle \boldsymbol{j}^2 \rangle$,
for both electrons (left panel) and ions (right panel).
The correlation between $Q_\omega$ and $Q_D$ is strong,
for both electrons and ions, indicating sheet-like vortex structures.
Meanwhile, this correlation for electrons is stronger than that for ions.
The correlation between $Q_\omega$ (or $Q_D$) and $Q_j$ is weak. }
\label{Fig.JpdfQWQDQJ}
\end{figure}

Several types of coherent structures emerge in turbulent flow, such as
$Q_D$ reflecting straining motions,
$Q_\omega$ corresponding to rotation and
$Q_j$ related to magnetic discontinuities.
All of them can interact with one another.
Fig. {\ref{Fig.JpdfQWQDQJ}}
shows the joint PDFs $P(x,y)$.


Generally, in many hydrodynamic turbulent flows,
the dominant structures are found to be
tube-like structures,
like vortex tubes with concentrated enstrophy (mean square vorticity),
whereas sheet-like structures between these tubes
are regions of dissipation.
It is therefore expected that
strain rate is not correlated with rotation rate.
In hydrodynamic turbulence
away from walls {\citep{Jimenez93, Blackburn96}},
the joint PDF of $Q_D$ versus $Q_\omega$ is spread
very broadly.
However, here
the joint PDFs $P(Q_\omega, Q_D)$ in Fig. {\ref{Fig.JpdfQWQDQJ}}
are dominated by a population near
the $Q_D=Q_\omega$ line,
which demonstrate the strong correlation
between these two quantities.
It also indicates
a frequently occurring class of sheet-like
rather than tube-like structures,
a feature consistent with many visualizations
of MHD turbulence {\citep{Dallas13}}.
We also find that the correlation between $Q_D$ and $Q_\omega$
for electrons (see Fig. {\ref{Fig.JpdfQWQDQJ}}(a.1))
is stronger than that for ions (see Fig. {\ref{Fig.JpdfQWQDQJ}}(a.2)).
Meanwhile, Fig. {\ref{Fig.ZoomCut}} confirms that the curves of $Q_D$ and $Q_\omega$
for electrons almost overlap,
while those for ions are more dispersed.

The joint PDFs of $Q_j$ versus $Q_D$ and $Q_\omega$,
shown in Fig. {\ref{Fig.JpdfQWQDQJ}},
are spread rather broadly,
indicating weak pointwise
correlation between these quantities.
By examining the curl of the Lorentz force,
one may deduce \citep{Matthaeus82,ParasharMatthaeus16}
that the vorticity structures tend to form on the
flanks of strong current structures.
Therefore, the vorticity distribution
does not correlate exactly with the current,
but are somewhat offset in space,
but nevertheless
located in the vicinity of the current.
This can be also observed in Fig. {\ref{Fig.ZoomCut}}.

\section{Energy transfer across scales}
\label{Sec.energy-transfer}
Now that the role of pressure tensor in
global energy conversion is established,
we turn to the
hierarchy of scales
and seek to find clues to
how energy cascade proceeds from
MHD scales down to the kinetic scales.
A low-pass spatial filter approach {\citep{Germano92}}
eliminates information at
small scales without affecting
remaining information at large scales,
thus providing
a powerful technique
to study cross-scale energy transfer.

Extension of the filtering approach to kinetic plasma
is technically easy. More derivation details
can be found in App. {\ref{appendsec:FKE}}.
Note that for conciseness of notation,
the filtering scale $\ell$
is not written explicitly unless necessary for clarity.
The equation of filtered fluid flow energy,
$\widetilde{E}^{f}_{\alpha}=\bar{\rho}_{\alpha} \tilde{\boldsymbol{u}}_{\alpha}^2/2$,
can be written as
\begin{equation}
\partial_t \widetilde{E}^{f}_{\alpha} + \nabla \cdot \boldsymbol{J}^u_{\alpha}
= -{\boldsymbol{\Pi}^{uu}_{\alpha}}-{\boldsymbol{\Phi}^{uT}_{\alpha}}-{\boldsymbol{\Lambda}^{ub}_{\alpha}}.
\label{Eq.FKE}
\end{equation}
The meaning of each term can be understood
referring to Eqs. {\ref{Eq.energy-conversion1}}-{\ref{Eq.energy-conversion3}} as:
\\
$\boldsymbol{J}^u_{\alpha}={\widetilde{E}^{f}_{\alpha} \tilde{\boldsymbol{u}}_{\alpha} + \bar{\rho}_{\alpha} \tilde{\boldsymbol{\tau}}^{u}_{\alpha} \cdot \tilde{\boldsymbol{u}}_{\alpha} + \overline{\boldsymbol{P}}_{\alpha} \cdot \tilde{\boldsymbol{u}}_{\alpha}}$
is the spatial transport;
\\
$\boldsymbol{\Pi}^{uu}_{\alpha}=-\left(\bar{\rho}_{\alpha} \tilde{\boldsymbol{\tau}}^{u}_{\alpha} \cdot \nabla\right) \cdot \tilde{\boldsymbol{u}}_{\alpha} - q_{\alpha} \bar{n}_{\alpha} \tilde{\boldsymbol{\tau}}^{b}_{\alpha} \cdot \tilde{\boldsymbol{u}}_{\alpha}$,
where $\tilde{\boldsymbol{\tau}}^{u}_{\alpha}= \left(\widetilde{\boldsymbol{u}_{\alpha} \boldsymbol{u}_{\alpha}} - \tilde{\boldsymbol{u}}_{\alpha} \tilde{\boldsymbol{u}}_{\alpha}\right)$
and  $\tilde{\boldsymbol{\tau}}^{b}_{\alpha}=\left(\widetilde{\boldsymbol{u}_{\alpha} \times \boldsymbol{B}}- \tilde{\boldsymbol{u}}_{\alpha} \times \widetilde{\boldsymbol{B}} \right)$,
is the flux of the fluid flow energy transfer across scales
(if the filter scale is set to zero, these terms vanish);
\\
$\boldsymbol{\Phi}^{uT}_{\alpha}=-\left(\overline{\boldsymbol{P}}_{\alpha} \cdot \nabla\right) \cdot \tilde{\boldsymbol{u}}_{\alpha}$
is the fluid flow energy converted into thermal (random) energy
due to pressure work;
\\
$\boldsymbol{\Lambda}^{ub}_{\alpha}=-q_{\alpha} \bar{n}_{\alpha} \widetilde{\boldsymbol{E}} \cdot \tilde{\boldsymbol{u}}_{\alpha}$
is the rate of fluid flow energy conversion into electromagnetic energy, i.e.,
electromagnetic work done on the fluid (seen more clearly in the filtered equation for electromagnetic energy in App. {\ref{appendsec:FME}}).

\subsection{Energy fluxes}
\begin{figure}[!htpb]
\centering
\includegraphics[width=0.45\textwidth]{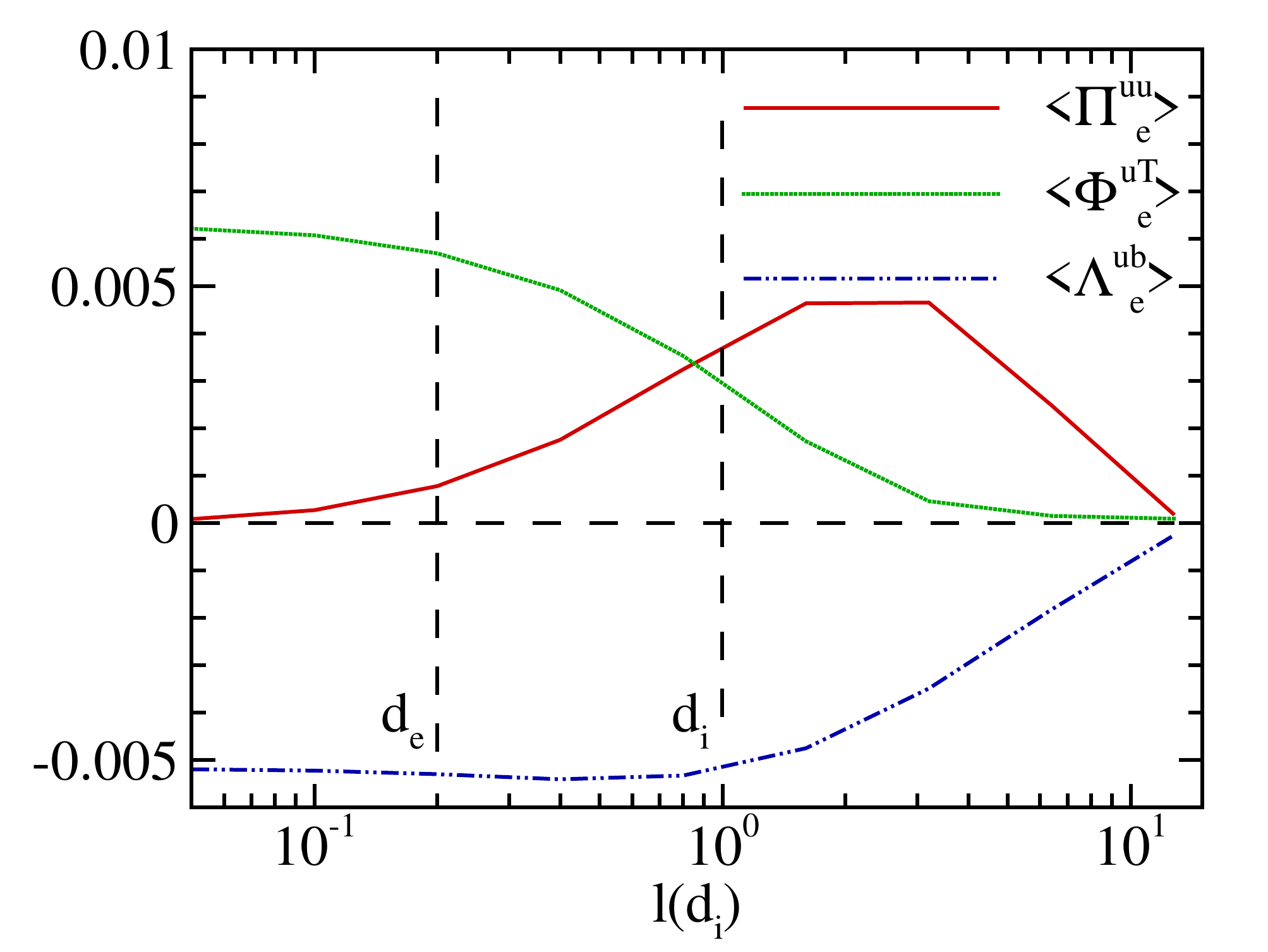}
\includegraphics[width=0.45\textwidth]{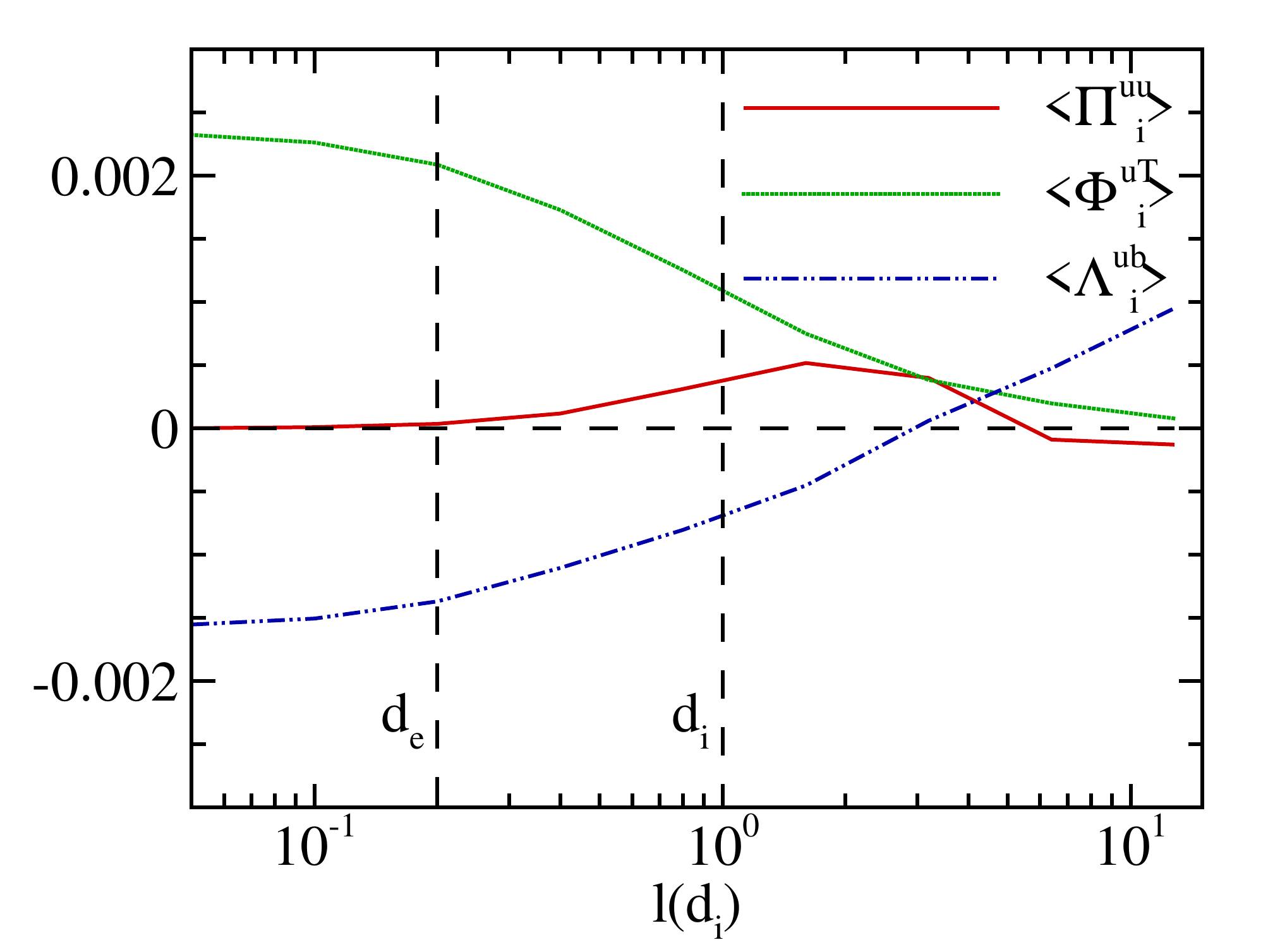}
\caption{Electron (left panel) and ion (right panel) energy transfer fluxes,
$\langle \boldsymbol{\Pi}^{uu}_{\alpha} \rangle$
(red solid line)
between small-scale and large-scale fluid flows,
$\langle \boldsymbol{\Phi}^{uT}_{\alpha} \rangle$
(green dashed line)
between fluid flow and random component, and
$\langle \boldsymbol{\Lambda}^{ub}_{\alpha} \rangle$
(blue dash-dotted line)
between fluid flow and magnetic field,
as a function of filtering length $\ell$.}
\label{Fig.filtersgsflux}
\end{figure}

We show energy fluxes,
$\langle \boldsymbol{\Pi}^{uu}_{\alpha} \rangle$,
$\langle \boldsymbol{\Phi}^{uT}_{\alpha} \rangle$
and $\langle \boldsymbol{\Lambda}^{ub}_{\alpha} \rangle$
varying with filtering scales
in Fig. {\ref{Fig.filtersgsflux}}.
The positive $\langle \boldsymbol{\Pi}^{uu}_{\alpha} \rangle$
transfers fluid flow energy from large to small scales
(i.e., a forward cascade) due to the interaction of
sub-grid scales with large scales, such as
the sub-scale stresses (SGS)
$\tilde{\boldsymbol{\tau}}^{u}_{\alpha}= \left(\widetilde{\boldsymbol{u}_{\alpha} \boldsymbol{u}_{\alpha}} - \tilde{\boldsymbol{u}}_{\alpha} \tilde{\boldsymbol{u}}_{\alpha}\right)$
and
$\tilde{\boldsymbol{\tau}}^{b}_{\alpha}=\left(\widetilde{\boldsymbol{u}_{\alpha} \times \boldsymbol{B}}- \tilde{\boldsymbol{u}}_{\alpha} \times \widetilde{\boldsymbol{B}} \right)$.
The flow energy cascade proceeds from the
largest scales
(here we calculate up to the correlation scale)
and transfer is sustained down to
electron inertial scale $d_e$.
The classical theory \cite{Kolmogorov41} in incompressible hydrodynamics
suggests an energy cascade
where energy is transferred from large to small scales
at a constant rate. This issue has been
studied in compressible HD and MHD, and
the SGS energy flux is found to be approximately constant;
see \cite{YangEA-PRE-16,Aluie11b,Aluie12,Wang13a}.
The idealized notion of
a constant cascade rate
is not applicable here,
likely due to
the limited scale separation, equivalent
to a relatively low Reynolds number.

Unlike the term $\langle \boldsymbol{\Pi}^{uu}_{\alpha} \rangle$
associated with interactions between
scales $>\ell$ and $<\ell$,
the terms $\langle \boldsymbol{\Phi}^{uT}_{\alpha} \rangle=\langle -\left(\overline{\boldsymbol{P}}_{\alpha} \cdot \nabla\right) \cdot \tilde{\boldsymbol{u}}_{\alpha} \rangle$
and $\langle \boldsymbol{\Lambda}^{ub}_{\alpha} \rangle=\langle -q_{\alpha} \bar{n}_{\alpha} \widetilde{\boldsymbol{E}} \cdot \tilde{\boldsymbol{u}}_{\alpha} \rangle$
incorporate information only
from scales $>\ell$.
Therefore, they are cumulative quantities.
Note that $\langle \boldsymbol{\Phi}^{uT}_{\alpha} \rangle$
is vanishingly small at large scales,
and increases as the filter scale approaching the small scales.
This indicates that the energy conversion
between fluid flow and random motion
by $\langle \boldsymbol{\Phi}^{uT}_{\alpha} \rangle$
is dominated by the contribution from
kinetic scales. Note that this term 
is the filtered version of the pressure dilatation
and {\it Pi-D} terms  discussed in Secs. \ref{Sec.GlobalEnergyConversion}
and \ref{Sec.RolePressureTensor}.

\begin{figure}[!htpb]
\centering
\begin{picture}(9,8)
\put(-0.2,0){\framebox(4.4,1){thermal (random) energy}}
\put(0,3){\framebox(9,2){}}\put(3,3.25){fluid flow energy}\put(7.5,3.5){\shortstack[r]{large\\scales}}
\put(0.5,3.5){\shortstack[l]{small\\scales}}
\put(5,7){\framebox(4,1){electromagnetic energy}}
\put(0.1,3){\color{green}\vector(0,-1){2}}\put(1,3){\color{green}\vector(0,-1){2}}\multiput(3,3)(1,0){2}{\color{green}\vector(0,-1){2}}
\put(2,3){\color{green}\line(0,-1){0.5}}\put(2,1.5){\color{green}\vector(0,-1){0.5}}\put(1.5,1.75){\color{green}$\langle \boldsymbol{\Phi}^{uT}_{\alpha} \rangle$}
\put(7.5,4){\color{red}\vector(-1,0){6}}
\put(4,4.25){\color{red}$\langle \boldsymbol{\Pi}^{uu}_{\alpha} \rangle$}
\multiput(5,7)(1,0){2}{\color{blue}\vector(0,-1){2}}\multiput(8,7)(1,0){2}{\color{blue}\vector(0,-1){2}}
\put(7,7){\color{blue}\line(0,-1){0.5}}\put(7,5.5){\color{blue}\vector(0,-1){0.5}}\put(6.5,5.75){\color{blue}$\langle \boldsymbol{\Lambda}^{ub}_{\alpha} \rangle$}
\end{picture}
\caption{Schematic diagram of energy transfer across spatial scales.
Electromagnetic energy is converted into both electron and ion fluid flows
at large scales due to electromagnetic work.
Pressure work converts these flows into random kinetic energies at small scales.
They are bridged by fluid flow energy transfer (turbulent spectral transfer)
across all spatial scales.}
\label{Fig.transdiagram}
\end{figure}
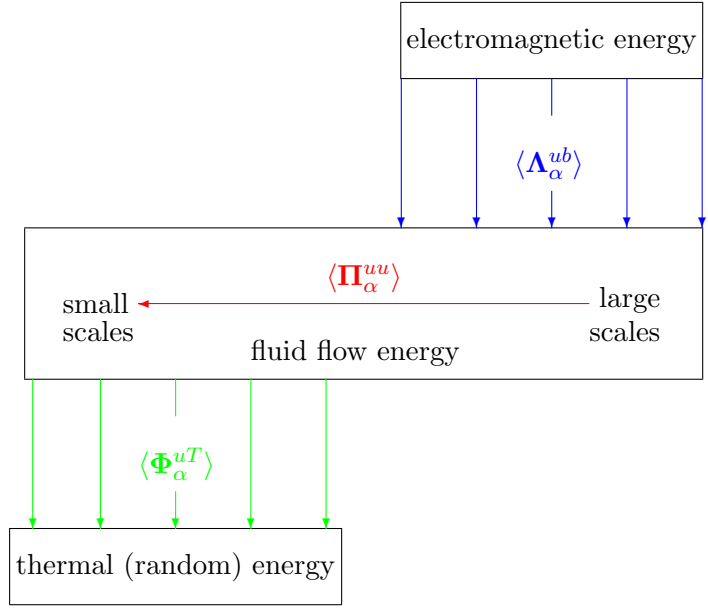

In contrast,
$\langle \boldsymbol{\Lambda}^{ub}_{\alpha} \rangle$,
the filtered contribution to 
$\langle -\boldsymbol{j}_\alpha \cdot \boldsymbol{E} \rangle$,
is fairly constant over kinetic scales,
and the observed
increases are concentrated at a few times
ion inertial scale $d_i$.
Therefore contributions to
the energy conversion between fluid flow
and electromagnetic fields
mainly result from the large scales in this simulation.
Simulation with larger size is required to
conclude if this transfer is macroscopic.
We schematically represent
this qualitative analysis
in Fig. {\ref{Fig.transdiagram}}.

\subsection{Location of energy transfer enhancement}
\begin{figure}[!htpb]
\centering
\includegraphics[width=0.9\textwidth]{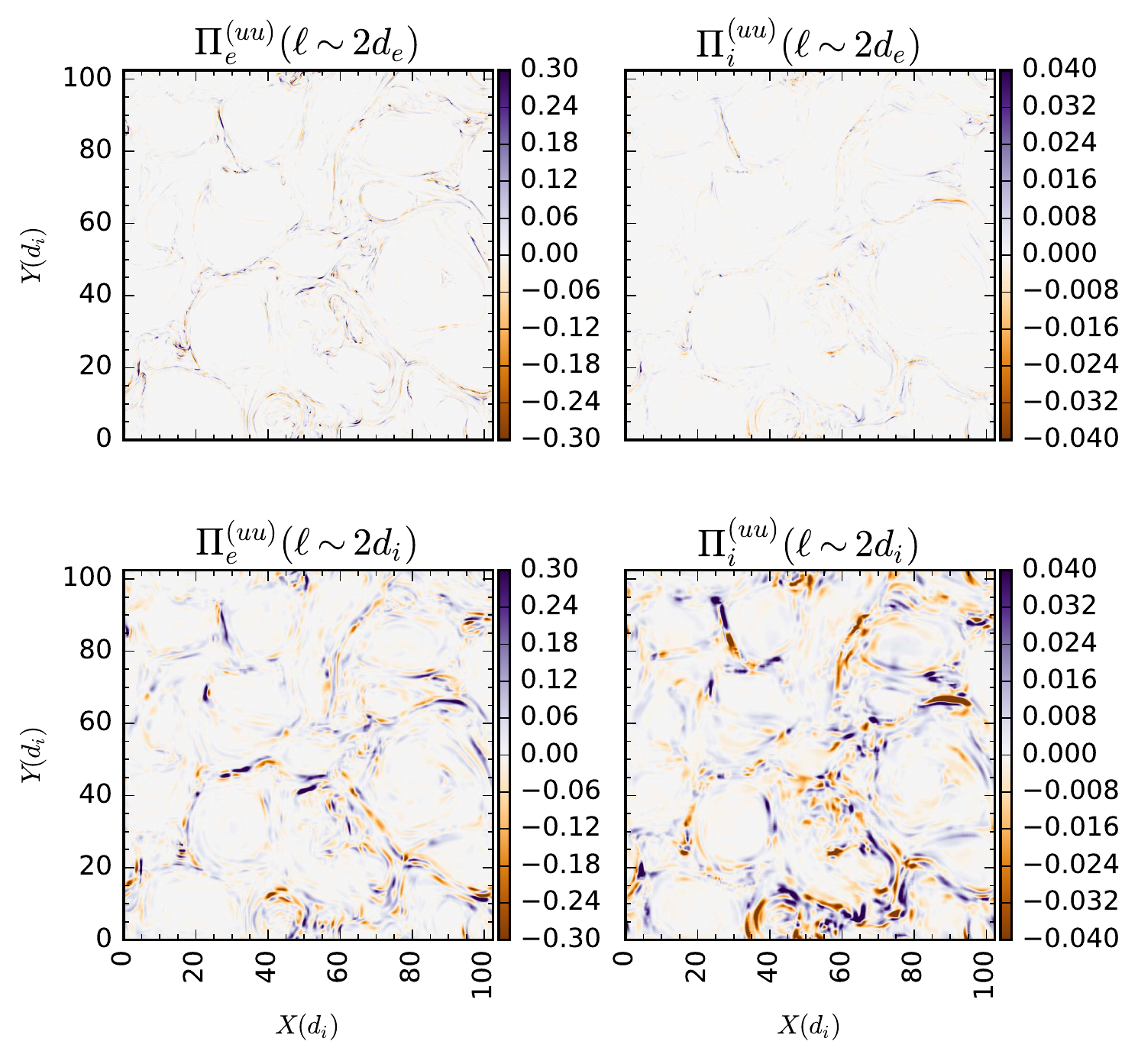}
\caption{Contours of species fluid flow energy transfer,
$\boldsymbol{\Pi}^{uu}_{\alpha}$,
across scales $\ell \sim 2d_e$ (top row)
and $\ell \sim 2d_i$ (bottom row)
for both electrons (left panel) and ions (right panel).}
\label{Fig.Contflux}
\end{figure}

In analogy with
the pressure work {\it Pi-D} in Fig. {\ref{Fig.ContPD}}(a)
and coherent structures in Figs. {\ref{Fig.ContQWQD}},
we portray
spatial contours of energy flux
$\boldsymbol{\Pi}^{uu}_{\alpha}$
in Fig. {\ref{Fig.Contflux}}. It is apparent that the
energy flux is highly inhomogeneous.
Comparing these figures,
it is apparent that
there is quite a significant
coincidence between the coherent structures and
the sites of enhanced energy transfer.
Arguments have been made
in favor of the hierarchy of
coherent structures (see {\citep{Karimabadi13}}).
It also appears clearly
in Fig. {\ref{Fig.Contflux}}.
In comparing the top row
with $\ell \sim 2d_e$
and the bottom row
with $\ell \sim 2d_i$,
the enhanced energy transfer spots
with $\ell \sim 2d_i$
are relatively broader.

\section{Conclusion}
Global energy equations derived from
the Vlasov-Maxwell system indicate the crucial role
of the pressure tensor in transforming
fluid flow energy into thermal (random) energy.
Accordingly for each species the electromagnetic
work ${\boldsymbol J} \cdot {\boldsymbol E}$, converts
energy between electromagnetic fields
and fluid flow energy, with a net transfer into
particle flows.
Finally the nonlinear transfer of energy across scales,
or the standard turbulence spectral transfer, provides a
more familiar coupling from large to small scales.
Taken together these quantities
provide a useful and essential vocabulary for
describing kinetic
turbulence and dissipation, and here we have provided
some basic information
about these processes based on
global, correlation and scale filtered
analysis of simulations.

The remarkable connection
between the pressure work $-\Pi_{ij}D_{ij}$ ({\it Pi-D})
and the normalized second invariant of
the traceless strain-rate tensor
$Q_D={\frac{1}{2}}D_{ij}D{ij}/ \langle 2D_{ij}D_{ij} \rangle$
is supported by the substantially similar patterns
of contour maps and cuts along the $X$ direction,
which corroborates again the idea suggested in Paper I.
In collisional fluid, $Q_D$ characterizes the viscous dissipation.
The energy transformation by $\langle -\Pi_{ij}D_{ij} \rangle$
in our case behaves like viscous dissipation.
We note that there is not any known quantitative theory
to explain this strong dependence rigorously
as far as we are aware. As
described, for example, by Del Sarto et al. {\citep{DelSartoPegoraro16}}
and Cerri et al. {\citep{Cerri14}},
velocity shear provides an effective mechanism
that can cause
an initially isotropic pressure to become nongyrotropic,
which can trigger a rich variety of inhomogeneous
instabilities and nonlinear effects {\cite{Servidio15}}. 
This might be of some help
to understand the result.
Further confirmation, based on
different codes, initial conditions, parameters and so on,
is needed to assert any degree of
universality of this result.
Once it is clarified, we might find
a pathway
(perhaps a
``viscous-like collisionless dissipation'')
by which various processes act in concert
and finally lead to intermittent heating.
It is noteworthy that this result is in some
ways complementary to the recent finding
\citep{WanEA15,WanEA16} that
in several types of kinetic simulations
of plasma turbulence,
conditional
averages of ${\boldsymbol j} \cdot {\boldsymbol E}$
based on local values of ${\boldsymbol j}$
behave very similarly to what would be expected if
${\boldsymbol j} \propto {\boldsymbol E}$, which is exact for MHD,
but does not emerge in any obvious way for a kinetic
plasma.
These results
represent a significant challenge to understand
based on theory, and also provide guidance for
developing
phenomenological theories of
plasma turbulence and dissipation.

Here we also
introduced several energy transfer
functions using
filtering approaches to
analyze their statistical
properties. This approach affords
a path to understand how
classical cascade theory is extended from MHD scales
down to kinetic scales.
One can envision that
energy exchange
between fluid flow and magnetic and electric fields
through $\langle \boldsymbol{E} \cdot \boldsymbol{j}_\alpha \rangle$
occurs at large scales, and
a part of fluid flow energy is converted into internal energy
through $\langle -\left( \boldsymbol{P}_\alpha \cdot \nabla \right) \cdot \boldsymbol{u}_\alpha \rangle$
at small scales. These two conversions are bridged
by the fluid flow energy cascade at moderate scales.

Another important aspect of plasma turbulence and energy
conversion that
we have described here is the association between
energy cascade, conversion processes
and coherent structures.
It transpires that
energy transfer and conversion are in general
inhomogeneous localized processes,
correlated with several types of structures,
such as velocity gradients
(i.e., symmetric straining and vortex) and current sheets.
Conversion into
random energy, in effect, dissipation,
is more strongly correlated with velocity gradients.
This situation is actually very complex,
involving a high correlation between
symmetric straining and vorticity, which requires special conditions
such as sheet-like structures.
Conversely
a weak correlation is found between current sheets
and vorticity (or symmetric straining),
even though these quantities are concentrated
in juxtaposed regions of space.
Clearly,
more work is required to reveal the dynamical
interactions among these coherent structures.
For this, we might gain some inspiration
from Refs. \citep{Matthaeus82, ParasharMatthaeus16} or
the evolution equation of enstrophy for
incompressible MHD,
\begin{equation}
d_t \left({\frac{1}{2}} \boldsymbol{\omega}^2\right) = \boldsymbol{\omega} \cdot \boldsymbol{D} \cdot \boldsymbol{\omega} + \nu \boldsymbol{\omega} \cdot \Delta \boldsymbol{\omega} + \boldsymbol{\omega} \cdot \nabla \times \left(\boldsymbol{j} \times \boldsymbol{B}\right)
\end{equation}

The 2.5D simulation used in this paper
presents some limitations with regard to potential
generality of the conclusions.
In fact
the present analysis
might miss some important effects present in 3D geometry
(in spite of recently reported qualitatively similarities
\cite{KarimabadiEASSR13,WanEA16,LiHowesEA16}).
The intent here is not to accommodate all the dynamical
complexity of the astrophysical plasma, but to
propose an alternative pathway
to study turbulence and heating in a
kinetic plasma.
Further development
of these ideas may be useful, e.g.,
in theoretical study of coronal heating and
acceleration of solar wind.
Finally,
neither the detailed differences
between electron and ion dynamics,
nor its implications for heating in space plasmas,
are discussed here, but
rather deferred for later study.

\begin{acknowledgments}
This research was partially
supported by AGS-1460130 (SHINE),
NASA grants NNX14AI63G (Heliophysics Grand Challenge Theory),
the Solar Probe Plus science team (ISOIS/SWRI subcontract No.
D99031L), and by the MMS Theory and Modeling team, NNX14AC39G.
We acknowledge
support provided by the China Scholarship Council
during a visit of Yan Yang to the University of Delaware.
\end{acknowledgments}

\appendix
\section{Filtered fluid flow energy equation}
\label{appendsec:FKE}

Consider the Vlasov equation,
\begin{equation}
\partial_t{f}_{\alpha}+\boldsymbol{v} \cdot \nabla{f_{\alpha}}+{\frac{q_{\alpha}}{m_{\alpha}}}\left(\boldsymbol{E}+\boldsymbol{v} \times \boldsymbol{B}\right) \cdot \nabla_{\boldsymbol{v}}{f_{\alpha}}=0.
\end{equation}
The spatially filtered $f_{\alpha}$ is given by
\begin{equation}
\bar{f}_{\alpha}\left(\boldsymbol{x},\boldsymbol{v},t\right)=\int{f_{\alpha}\left(\boldsymbol{x}',\boldsymbol{v},t\right)G_l(\boldsymbol{x}-\boldsymbol{x}') d\boldsymbol{x}'},
\end{equation}
where
$G_{\ell}\left(\boldsymbol{r}\right)={\ell}^{-3}G\left(\boldsymbol{r}/{\ell}\right)$
is a filtering kernel and
$G\left(\boldsymbol{r}\right)$ is a normalized boxcar
window function.
The low-pass filtered $\bar{f}_{\alpha}$
only contains information
at length scales $>\ell$.
The filtering operation can commute with derivative operations,
i.e.,
\begin{equation}
\overline{\partial_t{f_{\alpha}}}=\partial_t{\bar{f}_{\alpha}},\ \ \overline{\nabla{f_{\alpha}}}=\nabla{\bar{f}_{\alpha}},\ \ \overline{\nabla_{\boldsymbol{v}}{f_{\alpha}}}=\nabla_{\boldsymbol{v}}{\bar{f}_{\alpha}}.
\end{equation}
Then the spatially filtered Vlasov equation is written as:
\begin{equation}
\partial_t{\bar{f}_{\alpha}}+\boldsymbol{v} \cdot \nabla{\bar{f}_{\alpha}}+{\frac{q_{\alpha}}{m_{\alpha}}}\nabla_{\boldsymbol{v}} \cdot \left(\overline{\boldsymbol{E}f_{\alpha}}+\boldsymbol{v} \times \overline{\boldsymbol{B}f_{\alpha}}\right)=0.
\label{Eq.FVlasov}
\end{equation}

From Eq. {\ref{Eq.FVlasov}}, moment equations for each species yields
\begin{eqnarray}
\partial_t {\bar{\rho}_{\alpha}} + \nabla \cdot \left(\overline{\rho_{\alpha} \boldsymbol{u}_{\alpha}}\right) &=& 0, \\
\partial_t {\left(\overline{\rho_{\alpha} \boldsymbol{u}_{\alpha}}\right)} + \nabla \cdot \left(\overline{\rho_{\alpha} \boldsymbol{u}_{\alpha}\boldsymbol{u}_{\alpha}}\right) &=&  - \nabla \cdot {\overline{\boldsymbol{P}}_{\alpha}} + q_{\alpha} \left(\overline{n_{\alpha} \boldsymbol{E}}+\overline{n_{\alpha} \boldsymbol{u}_{\alpha} \times \boldsymbol{B}}\right).
\end{eqnarray}
One can see that these equations can also
be derived immediately from the macroscopic
Eqs. {\ref{Eq.1}} and {\ref{Eq.2}}.
A Favre-filtered (density-weighted-filtered)
field {\citep{Favre69}} is defined as
\begin{equation}
\tilde{\boldsymbol{a}}=\overline{n\boldsymbol{a}}/\bar{n}.
\end{equation}
Then the moment equations aforementioned can be written as
\begin{eqnarray}
\partial_t {\bar{\rho}_{\alpha}} + \nabla \cdot \left(\bar{\rho}_{\alpha} \tilde{\boldsymbol{u}}_{\alpha}\right) &=& 0, \\
\partial_t {\left(\bar{\rho}_{\alpha} \tilde{\boldsymbol{u}}_{\alpha}\right)} + \nabla \cdot \left(\bar{\rho}_{\alpha} \tilde{\boldsymbol{u}}_{\alpha} \tilde{\boldsymbol{u}}_{\alpha}\right) &=&  - \nabla \cdot {(\bar{\rho}_{\alpha} \tilde{\boldsymbol{\tau}}^{u}_{\alpha})} - \nabla \cdot {\overline{\boldsymbol{P}}_{\alpha}} + q_{\alpha} \bar{n}_{\alpha} \left(\widetilde{\boldsymbol{E}}+ \widetilde{\boldsymbol{u}_{\alpha} \times \boldsymbol{B}}\right),
\label{Eq.FMomentum}
\end{eqnarray}
where $\tilde{\boldsymbol{\tau}}^{u}_{\alpha}= \left(\widetilde{\boldsymbol{u}_{\alpha} \boldsymbol{u}_{\alpha}} - \tilde{\boldsymbol{u}}_{\alpha} \tilde{\boldsymbol{u}}_{\alpha}\right)$.
Eq. {\ref{Eq.FMomentum}} dot product
$\tilde{\boldsymbol{u}}_{\alpha}$,
we can get the equation of filtered fluid flow energy
$\widetilde{E}^{f}_{\alpha}=\bar{\rho}_{\alpha} |\tilde{\boldsymbol{u}}_{\alpha}|^2/2$,
\begin{eqnarray}
\partial_t \widetilde{E}^{f}_{\alpha} + \nabla \cdot \left( {\widetilde{E}^{f}_{\alpha} \tilde{\boldsymbol{u}}_{\alpha} + \bar{\rho}_{\alpha} \tilde{\boldsymbol{\tau}}^{u}_{\alpha} \cdot \tilde{\boldsymbol{u}}_{\alpha} + \overline{\boldsymbol{P}}_{\alpha} \cdot \tilde{\boldsymbol{u}}_{\alpha}} \right)
&=&\left(\bar{\rho}_{\alpha} \tilde{\boldsymbol{\tau}}^{u}_{\alpha} \cdot \nabla\right) \cdot \tilde{\boldsymbol{u}}_{\alpha} +q_{\alpha} \bar{n}_{\alpha} \tilde{\boldsymbol{\tau}}^{b}_{\alpha} \cdot \tilde{\boldsymbol{u}}_{\alpha}+ \nonumber \\
&& \left(\overline{\boldsymbol{P}}_{\alpha} \cdot \nabla\right) \cdot \tilde{\boldsymbol{u}}_{\alpha} + q_{\alpha} \bar{n}_{\alpha} \widetilde{\boldsymbol{E}} \cdot \tilde{\boldsymbol{u}}_{\alpha},
\label{Eq.Fkinetic}
\end{eqnarray}
where $\tilde{\boldsymbol{\tau}}^{b}_{\alpha}=\left(\widetilde{\boldsymbol{u}_{\alpha} \times \boldsymbol{B}}- \tilde{\boldsymbol{u}}_{\alpha} \times \widetilde{\boldsymbol{B}} \right)$.
We can use simple notations as defined in the text following Eq. \ref{Eq.FKE}:
\\
$\boldsymbol{J}^u_{\alpha}={\widetilde{E}^{f}_{\alpha} \tilde{\boldsymbol{u}}_{\alpha} + \bar{\rho}_{\alpha} \tilde{\boldsymbol{\tau}}^{u}_{\alpha} \cdot \tilde{\boldsymbol{u}}_{\alpha} + \overline{\boldsymbol{P}}_{\alpha} \cdot \tilde{\boldsymbol{u}}_{\alpha}}$
is the spatial transport;
\\
$\boldsymbol{\Pi}^{uu}_{\alpha}=-\left(\bar{\rho}_{\alpha} \tilde{\boldsymbol{\tau}}^{u}_{\alpha} \cdot \nabla\right) \cdot \tilde{\boldsymbol{u}}_{\alpha} - q_{\alpha} \bar{n}_{\alpha} \tilde{\boldsymbol{\tau}}^{b}_{\alpha} \cdot \tilde{\boldsymbol{u}}_{\alpha}$
is the flux of large-scale fluid flow energy transferred to sub-scale fluid flow energy;
\\
$\boldsymbol{\Phi}^{uT}_{\alpha}=-\left(\overline{\boldsymbol{P}}_{\alpha} \cdot \nabla\right) \cdot \tilde{\boldsymbol{u}}_{\alpha}$
is the rate of fluid flow energy converted into thermal (random) energy;
\\
$\boldsymbol{\Lambda}^{ub}_{\alpha}=-q_{\alpha} \bar{n}_{\alpha} \widetilde{\boldsymbol{E}} \cdot \tilde{\boldsymbol{u}}_{\alpha}$
is the rate of fluid flow energy converted into electromagnetic energy.

\section{Filtered electromagnetic energy equation}
\label{appendsec:FME}
From Maxwell's equations, we can obtain the equation for filtered electromagnetic energy
$\overline{E}^m={\frac{1}{8\pi}}\left(|\overline{\boldsymbol{B}}|^2+|\overline{\boldsymbol{E}}|^2\right)$,
\begin{equation}
\partial_t \overline{E}^m + {\frac{c}{4\pi}}\nabla \cdot \left(\overline{\boldsymbol{E}} \times \overline{\boldsymbol{B}}\right)= - \overline{\boldsymbol{E}} \cdot \bar{\boldsymbol{j}},
\label{Eq.FME}
\end{equation}
Taking the electromagnetic work (i.e., $\boldsymbol{\Lambda}^{ub}_{\alpha}$ in Eqs. \ref{Eq.FKE} and \ref{Eq.Fkinetic}) as a source
in Eq. \ref{Eq.FME} yields
\begin{equation}
\partial_t \overline{E}^m + {\frac{c}{4\pi}}\nabla \cdot \left(\overline{\boldsymbol{E}} \times \overline{\boldsymbol{B}}\right)= \sum_\alpha{q_\alpha \bar{n}_\alpha \left(\widetilde{\boldsymbol{E}}-\overline{\boldsymbol{E}}\right) \cdot \tilde{\boldsymbol{u}}_\alpha} - \sum_\alpha{q_\alpha \bar{n}_\alpha \widetilde{\boldsymbol{E}} \cdot \tilde{\boldsymbol{u}}_\alpha},
\end{equation}
which can be written as
\begin{equation}
\partial_t \overline{E}^m + \nabla \cdot \boldsymbol{J}^b = -\sum_\alpha {\boldsymbol{\Pi}^{bb}_{\alpha}} + \sum_\alpha {\boldsymbol{\Lambda}^{ub}_{\alpha}},
\end{equation}
where
\\
$\boldsymbol{J}^b={\frac{c}{4\pi}} \left(\overline{\boldsymbol{E}} \times \overline{\boldsymbol{B}}\right)$ is the spatial transport;
\\
$\boldsymbol{\Pi}^{bb}_{\alpha}=-q_\alpha \bar{n}_\alpha  \tilde{\boldsymbol{\tau}}^{e}_{\alpha} \cdot \tilde{\boldsymbol{u}}_\alpha$,
where $\tilde{\boldsymbol{\tau}}^{e}_{\alpha}=\left(\widetilde{\boldsymbol{E}}-\overline{\boldsymbol{E}}\right)$,
is the flux of electromagnetic energy across scales due to sub-scale work done by the electric field;
\\
$\boldsymbol{\Lambda}^{ub}_{\alpha}=-q_{\alpha} \bar{n}_{\alpha} \widetilde{\boldsymbol{E}} \cdot \tilde{\boldsymbol{u}}_{\alpha}$,
the rate of fluid flow energy conversion into electromagnetic energy, is the same as those in Eqs. \ref{Eq.FKE} and \ref{Eq.Fkinetic}, which is canceled out
on combining Eqs. \ref{Eq.FKE} (or \ref{Eq.Fkinetic}) and \ref{Eq.FME}. Therefore, the filtered equation for total fluid flow and electromagnetic energy
takes the form
\begin{equation}
\partial_t \left(\sum_\alpha \widetilde{E}^{f}_{\alpha} + \overline{E}^m\right) + \nabla \cdot \left(\sum_\alpha \boldsymbol{J}^u_{\alpha} + \boldsymbol{J}^b\right)
= -\sum_\alpha {\boldsymbol{\Pi}^{uu}_{\alpha}} - \sum_\alpha {\boldsymbol{\Pi}^{bb}_{\alpha}} - \sum_\alpha {\boldsymbol{\Phi}^{uT}_{\alpha}},
\end{equation}
where
\\
$\boldsymbol{J}^u_{\alpha}={\widetilde{E}^{f}_{\alpha} \tilde{\boldsymbol{u}}_{\alpha} + \bar{\rho}_{\alpha} \tilde{\boldsymbol{\tau}}^{u}_{\alpha} \cdot \tilde{\boldsymbol{u}}_{\alpha} + \overline{\boldsymbol{P}}_{\alpha} \cdot \tilde{\boldsymbol{u}}_{\alpha}}$,
\\
$\boldsymbol{\Pi}^{uu}_{\alpha}=-\left(\bar{\rho}_{\alpha} \tilde{\boldsymbol{\tau}}^{u}_{\alpha} \cdot \nabla\right) \cdot \tilde{\boldsymbol{u}}_{\alpha} - q_{\alpha} \bar{n}_{\alpha} \tilde{\boldsymbol{\tau}}^{b}_{\alpha} \cdot \tilde{\boldsymbol{u}}_{\alpha}$ and 
\\
$\boldsymbol{\Phi}^{uT}_{\alpha}=-\left(\overline{\boldsymbol{P}}_{\alpha} \cdot \nabla\right) \cdot \tilde{\boldsymbol{u}}_{\alpha}$
are defined in the text following Eqs. \ref{Eq.FKE} and \ref{Eq.Fkinetic}. 

 \newcommand{\BIBand} {and} 
  \newcommand{\boldVol}[1] {\textbf{#1}} 
  \providecommand{\SortNoop}[1]{} 
  \providecommand{\sortnoop}[1]{} 
  \newcommand{\stereo} {\emph{{S}{T}{E}{R}{E}{O}}} 
  \newcommand{\au} {{A}{U}\ } 
  \newcommand{\AU} {{A}{U}\ } 
  \newcommand{\MHD} {{M}{H}{D}\ } 
  \newcommand{\mhd} {{M}{H}{D}\ } 
  \newcommand{\RMHD} {{R}{M}{H}{D}\ } 
  \newcommand{\rmhd} {{R}{M}{H}{D}\ } 
  \newcommand{\wkb} {{W}{K}{B}\ } 
  \newcommand{\alfven} {{A}lfv{\'e}n\ } 
  \newcommand{\alfvenic} {{A}lfv{\'e}nic\ } 
  \newcommand{\Alfven} {{A}lfv{\'e}n\ } 
  \newcommand{\Alfvenic} {{A}lfv{\'e}nic\ }

\end{document}